\documentclass[twocolumn]{aastex631}
\usepackage{appendix}
\usepackage{graphicx}

\usepackage{CJKutf8}
\usepackage{tabularx}
\usepackage{lineno}

\shorttitle{A catalog of candidate WD+MS binaries in open clusters}
\shortauthors{Grondin et al.}

\begin{document}

\title{The first catalog of candidate white dwarf-main sequence binaries \\ in open star clusters: A new window into common envelope evolution}

\correspondingauthor{Steffani M. Grondin}
\email{steffani.grondin@astro.utoronto.ca}

\author[0000-0002-0444-8502]{Steffani M. Grondin}
\affiliation{David A. Dunlap Department of Astronomy \& Astrophysics, University of Toronto,
50 St. George St., Toronto, ON M5S 3H4, Canada }

\author[0000-0001-7081-0082]{Maria R. Drout}
\affiliation{David A. Dunlap Department of Astronomy \& Astrophysics, University of Toronto,
50 St. George St., Toronto, ON M5S 3H4, Canada }

\author[0000-0002-5608-4683]{Jason Nordhaus}
\affiliation{National Technical Institute for the Deaf, Rochester Institute of Technology, 1 Lomb Memorial Dr., Rochester, NY 14623, USA}
\affiliation{Center for Computational Relativity and Gravitation, Rochester Institute of Technology, \\ 1 Lomb Memorial Dr.,  Rochester, NY 14623, USA}

\author[0000-0002-0638-8822]{Philip S. Muirhead}
\affiliation{Department of Astronomy \& The Institute for Astrophysical Research, Boston University, \\ 725 Commonwealth Ave., Boston, MA 02215,
USA}

\author[0000-0003-2573-9832]{Joshua S. Speagle (\begin{CJK*}{UTF8}{gbsn}沈佳士\ignorespacesafterend\end{CJK*})}
\affiliation{Department of Statistical Sciences, University of Toronto, 9th Floor, Ontario Power Building, \\ 700 University Ave., Toronto, ON M5S 3G3, Canada}
\affiliation{David A. Dunlap Department of Astronomy \& Astrophysics, University of Toronto, 
50 St. George St., Toronto, ON M5S 3H4, Canada }
\affiliation{Dunlap Institute for Astronomy and Astrophysics, University of Toronto, 50 St George St., Toronto, ON M5S 3H4, Canada}
\affiliation{Data Sciences Institute, University of Toronto, 17th Floor, Ontario Power Building,
\\ 700 University Ave., Toronto, ON M5G 1Z5, Canada}

\author[0000-0002-7706-5668]{Ryan Chornock}
\affiliation{Department of Astronomy, University of California, Berkeley, CA 94720-3411, USA}

\received{July 5, 2024}
\accepted{for publication in The Astrophysical Journal on August 15, 2024}

\begin{abstract}
Close binary systems are the progenitors to both Type Ia supernovae and the compact object mergers that can be detected via gravitational waves. To achieve a binary with a small radial separation, it is believed that the system likely undergoes common envelope (CE) evolution. Despite its importance, CE evolution may be one of the 
largest uncertainties in binary evolution due to a combination of computational challenges and a lack of observed benchmarks where both the post-CE and pre-CE conditions are known. Identifying post-CE systems in star clusters can partially circumvent this second issue by providing an independent age constraint on the system. For the first time, we conduct a systematic search for white dwarf (WD) and main-sequence (MS) binary systems in 299 Milky Way open star clusters. Coupling \textit{Gaia} DR3 photometry and kinematics with multi-band photometry from Pan-STARRS1 and 2MASS, we apply a machine learning based approach and find 52 high-probability candidates in 38 open clusters. For a subset of our systems, we present follow-up spectroscopy from the Gemini and Lick Observatories and archival light curves from \textit{TESS}, \textit{Kepler/K2} and the Zwicky Transient Facility. Examples of M-dwarfs with hot companions are spectroscopically observed, along with regular system variability. While the kinematics of our candidates are consistent with their host clusters, some systems have spatial positions offset relative to their hosts, potentially indicative of natal kicks. Ultimately, this catalog is a first step to obtaining a set of observational benchmarks to better link post-CE systems to their pre-CE progenitors. \\
\end{abstract}

\vspace{-0.5in}

\section{Introduction} \label{sec:intro}

White dwarfs (WDs) are the final evolutionary outcome for nearly all stars in the Universe. On their own, WDs are impressive cosmic laboratories, informing stellar evolution histories through the initial-final mass relation \citep[IFMR, e.g.][]{2018ApJ...866...21C} and equations of state of degenerate matter \citep[][]{1990RPPh...53..837K}. WDs in binary systems are especially useful, since they are the progenitors to both Type Ia supernovae and mergers producing gravitational waves. Fortunately, binaries containing a WD are relatively common, where $\sim$$25\%$ of binaries in the immediate solar vicinity ($<20$ parsecs) have a companion \citep{2008AJ....135.1225H, 2017A&A...602A..16T}. Unfortunately, we often lack context for how a binary system evolved into its current configuration, and binary evolution models rely on multiple uncertain physical processes, from the stability and efficiency of mass transfer to the impact of natal kicks. These uncertainties then impact multiple areas of astrophysics, from predictions for compact object merger rates \citep{2013A&ARv..21...59I} and gravitational wave detections \citep{2017MNRAS.471.4702B} to our understanding of planetary survival during single star evolution \citep{1998AJ....116.1308S, 1998A&A...335L..85N, 2013MNRAS.432..500N, 2021MNRAS.501..676L, 2021ApJ...915L..34M}.

The stability of mass-transfer primarily depends on two quantities: (i) how the radius of the primary star responds to mass loss and (ii) how the orbit of the system itself responds to the mass transfer \citep[e.g.][]{1993PASP..105.1373I, pols2011stellar}. Stable mass-transfer will occur if the transfer causes the primary star to either contract more or expand less than its Roche lobe. However, if the response to mass transfer leads to a situation where the Roche radius is located inside the nominal surface of the primary star (often the case for red giant or asymptotic giant branch stars with deep convective envelopes that can expand in response to mass loss) mass will be unstably transferred to the companion star on a dynamical timescale.
In this case, the transferred mass collects in an extended envelope around both stars, causing the system to orbit inside a \textit{common envelope} \citep[CE, e.g.][]{1988ApJ...329..764L, 2008ASSL..352..233W, 2013Sci...339..433I, 2013A&ARv..21...59I, 2014LRR....17....3P}. Once inside the CE, the binary's orbit loses energy and angular momentum. As a result, the binary either ejects the envelope, emerging as a short-period post-CE system, or is tidally disrupted and merges, leaving a single star whose evolution has been substantially altered \citep{2011PNAS..108.3135N, 2019MNRAS.490.1179G, 2022MNRAS.511.5994G}.

Understanding the CE phase is crucial for determining the evolution of short-period binary systems \citep{1976IAUS...73...75P}. Unfortunately, modelling CE evolution is extremely difficult due to the large range of  scales and complex physical processes involved. As a result, population synthesis models often rely on heuristic prescriptions and approximations of important quantities like the envelope ejection efficiency parameter \citep[`$\alpha_{CE}$', e.g.][]{1984ApJ...277..355W} to approximate CE outcomes. While there has been a large effort to conduct hydrodynamical simulations to better understand the CE phase, comparison of the outcomes is difficult given the large array of input physical quantities \citep[for a recent comparison of 3D CE simulations to observations in the context of $\alpha_{CE}$, see][]{2019MNRAS.490.2550I}. Often, these simulations result in predictions of binary outcomes inconsistent with observations of post-CE field systems \citep{2012ApJ...746...74R, 2012ApJ...744...52P, 2021A&A...648L...6P} though the inclusion of missing physical processes such as convection and radiation may resolve this issue \citep{2019MNRAS.485.4492W, 2020MNRAS.497.1895W, 2022MNRAS.516.2189W}. Even so, it is clear that observational constraints of \textit{both the pre- and post-CE phases} are essential.

The renaissance of large multi-wavelength surveys has yielded the discovery of thousands of Galactic binary systems containing a WD and main-sequence (MS) star \citep{2004A&A...419.1057W,2006ApJ...646..480F, 2010MNRAS.402..620R, 2014A&A...570A.107R,2016MNRAS.463.2125P,2021MNRAS.504.2420I, 2021MNRAS.506.5201R, 2023MNRAS.tmp.3454N}, and a few dozen post-CE WD+MS binaries have been identified in this data \citep[e.g.][]{2010MNRAS.402..620R, 2011A&A...536A..43N, 2012MNRAS.424.1841L, 2013MNRAS.432..500N,2021MNRAS.504.2420I, 2022MNRAS.517.2867H}. While this sample of systems allows for understandings of post-CE behaviour, they are located in the field and thus suffer from cooling track uncertainties due to their binary nature. 

Identifying WD+MS post-CE systems in stellar clusters can circumvent this issue by providing independent constraints on the age of the system. This, in turn, allows for more detailed probes of the pre-CE evolution. Such efforts have been successfully applied to two post-CE WD+MS binaries in the Hyades: V471 Tau and HZ9. For V471 Tau, \cite{2022AJ....163...34M} estimate the pre-CE mass of the WD progenitor to be between $M=3.16 M_\odot - 3.41 M_{\odot}$. Combined with the WD's temperature and age of the Hyades, this progenitor mass implies that the system could not have formed via standard binary evolution. Both \cite{2022AJ....163...34M} and \cite{2001ApJ...563..971O} favor a scenario where a merger occurred in a triple system and subsequently entered a CE with the tertiary star after the merged product evolved off the MS. In the case of HZ9, Muirhead et al., in prep measure a current mass for the WD of $M=0.504 \pm 0.016 M_{\odot}$. Combined with the cooling age, this is compatible with isolated binary evolution for a system that underwent a CE phase.

Despite their utility, only a handful of WD+MS systems have ever been associated with an open cluster (OC), likely because 5-D kinematic (position, parallax and proper motion) information of stars in the Galactic disk was scarce prior to the launch of \textit{Gaia} \citep[for a recent review on binary system insights made possible by \textit{Gaia}, see][]{2024arXiv240312146E}. To the best of our knowledge, the only WD+MS binaries found in OCs are: (i) V471 Tau (ii) HZ9 and (iii) a magnetized WD+MS binary associated with NGC 2422 \citep{2019ApJ...880...75R}. In the latter case, no orbital period has yet been measured, so it is unclear whether it is indeed a post-CE system. In addition, this system's faint \textit{Gaia} apparent magnitude ($m_{G}=19.8$ mag) makes additional follow-up challenging. Clearly, additional benchmark post-CE WD+MS binaries in stellar clusters would be beneficial.

In this work, we perform the first systematic search for WD+MS binary systems in 299 Milky Way OCs. We define our OC and stellar samples in Section \ref{sec:search}, and outline the supervised machine learning approach used to photometrically identify WD+MS binary candidates in this study in Section \ref{sec:initcand}. Section \ref{sec:photocand} presents the photometric, spatial and kinematic properties of our high-probability WD+MS candidates, while follow-up spectra and archival light curves for an example subset of these systems are presented in Section \ref{sec:verification}. In Section \ref{sec:contaminants}, we discuss other possible astrophysical sources present in our catalog by photometrically comparing our candidates to known samples of cataclysmic variables, MS+MS binaries, rapidly rotating (active) M-dwarfs and single WDs. We discuss implications of natal kicks in Section \ref{sec:natal} and summarize in Section \ref{sec:summary}. Spectral energy distributions (SEDs) for all of our candidates are presented in Appendix \ref{sec:seds}. Ultimately, this catalog is an essential first step in identifying a population of WD+MS post-CE binaries in clusters.

\section{Initial Sample Selection} \label{sec:search}

In this paper, we aim to photometrically identify candidate WD+MS binaries in stellar clusters based on multi-band photometry. Here, we describe both our choice of which Milky Way OCs to search within (Section \ref{sec:openclusters}) and which specific stars in the vicinity of those clusters we will further examine below (Section \ref{sec:starsample}).

\subsection{Open Cluster Selection} \label{sec:openclusters}

Our sample of Milky Way OCs stems from the \cite{2020A&A...640A...1C} catalog -- a compilation of 2017 OCs identified in \textit{Gaia} DR2 \citep{2016A&A...595A...1G, 2018A&A...616A...1G}. Since we are searching for WD+MS systems, the OCs in our study must adhere to specific distance, declination and age criteria. First, we discard 150 OCs in this sample that do not have reported distances or ages. The rest of our selection criteria to produce our final OC sample is described below.  

\subsubsection{Distance}

We place an upper limit on the distance to the clusters that we search in this study to ensure that (i) WD+MS binaries are detectable given the depth of the surveys we utilize and (ii) reliable \textit{Gaia} parallaxes are expected to be broadly available. To assess the detectability of WD+MS binaries, we consider a known sample of spectroscopic WD+MS systems from the Sloan Digital Sky Survey \citep[SDSS,][]{2010MNRAS.402..620R}. While there are some bright outliers ($M_{G} \sim 6$ mag), these binaries generally have absolute magnitudes between $M_{G} = 8-14$ mag in the \textit{Gaia} G-band. At distances $\gtrsim 1.5$ kpc, bright WD+MS binaries would have magnitudes fainter than $m_{G}=19$ mag. Moreover, \textit{Gaia} DR3 kinematics are essential for associating binaries with host clusters in this study, so it is imperative that distance measurements to the cluster stars are reliable. At large distances, using inverted parallax as a distance estimate is challenging due to the resulting skewed distance error distribution \citep{1973PASP...85..573L}. Depending on color and sky position, sources at distances further than $\sim 1.5$ kpc have errors in parallax $\gtrsim 20\%$ and are thus less reliable \citep{BailerJones2015, 2021AJ....161..147B}. Hence, we restrict ourselves to clusters at distances $<1.5$ kpc, as at distances greater than this, (i) even bright WD+MS binaries become harder to detect and (ii) parallax measurements become less reliable. Imposing this constraint on the full OC sample in \cite{2020A&A...640A...1C} yields 589 clusters.

\subsubsection{Declination}

As we are \textit{photometrically} identifying WD+MS binaries, we utilize multi-band data from a variety of astronomical surveys in this study. The presence of a WD can be inferred by a photometric excess at blue wavelengths \citep[e.g.][]{2010MNRAS.402..620R}, so we impose a minimum constraint that each candidate must have at least $g$-band data. The Panoramic Survey Telescope and Rapid Response System \citep[Pan-STARRS1,][]{2012ApJ...750...99T, 2016arXiv161205560C} has collected $g$, $r$, $i$, and $z$-band photometry for much of the sky with declinations $> -30^{\circ}$. While there are indeed other surveys that offer $g$-band magnitudes (e.g.\ Skymapper; \citealt{Onken2024}), we restrict ourselves to clusters that lie within the Pan-STARRS1 footprint for uniformity across photometric bands. When we impose this filter, we are left with 403 clusters in our sample.

\subsubsection{Age}

Finally, as our goal is to identify candidate WD+MS binaries, we limit ourselves to clusters old enough such that the first, most massive, WDs will have had time to form. While the precise divide between high and low mass stellar evolution is debated, the divide between WD and neutron star (or black hole) formation is expected at masses $\sim 8 \mathrm{M_{\odot}}$ \citep{2008ApJ...675..614P}. Thus, to determine an approximate minimum age for a cluster to contain WDs, we examine when a $7 \mathrm{M_\odot}$ MS star reaches the tip of the thermally pulsing asymptotic giant branch with the MESA Isochrones and Stellar Tracks \citep[\texttt{MIST,}][]{2016ApJS..222....8D, 2016ApJ...823..102C}. We find that it takes $\sim 55$ Myr for this star to become a WD, so we impose a minimum age constraint by only searching for WD+MS systems in clusters that have ages $> 50$ Myr. \cite{2020A&A...640A...1C} measure cluster ages via a neural network, where \textit{301 clusters that passed our previous cuts also have ages larger than 50 Myr and thus represent our final sample of OCs to search in this study.}

\subsection{Stellar Sample Selection}  \label{sec:starsample}
For each of the 301 OCs identified in Section \ref{sec:openclusters}, we utilize \textit{Gaia} DR3 data \citep{2016A&A...595A...1G, 2023A&A...674A...1G} to define the set of stars that we further analyze in Section \ref{sec:initcand} to identify possible WD+MS systems. Each sample of stars is selected by combining (i) a generous search radius around each cluster and kinematic constraints to identify stars broadly consistent with cluster membership, (ii) photometric constraints to select stars in the broad region of the \textit{Gaia} color-magnitude diagram where WD+MS systems are expected to appear, and  (iii) the requirement that sufficient multi-band photometry is available to carry out a most detailed assessment of each system's possible binary nature. Here, we describe each of these constraints in more detail. With respect to point (i) we note that we are purposely generous during our initial sample selection, while a more detailed assessment of the kinematics of high-probability WD+MS systems is performed in Section \ref{sec:kinematics-candidates}.

\subsubsection{Cluster Search Radius} \label{sec:searchradius}

When identifying members of an OC, a physical radius of 50 parsecs from the cluster centre is often chosen as a baseline search radius \citep[e.g.][]{2023arXiv230308474V}. However, recent observations show that many OCs are surrounded by extended coronae of stars \citep{2021A&A...645A..84M, 2022A&A...659A..59T, 2024arXiv240618767K}. Moreover, post-CE WD+MS binaries could receive natal kicks of a few km s$^{-1}$ during asymmetric mass loss in the asymptotic giant branch phase \citep[e.g.][]{1993ApJ...413..641V, 2007MNRAS.381L..70H} or upon ejection of the CE \citep{1998ApJ...500..909S, 2019MNRAS.486.1070C}. As such, WD+MS binaries could end up on the cluster outskirts or be located at larger radii than most known cluster members.

Thus, to encapsulate as many WD+MS binaries as possible---including those that could be post-CE---while still selecting sources in close proximity to each OC, we select the search radius for each OC via a two-step process. First, we compute the physical distance, $r_{prob}$, corresponding to a location $30\%$ larger than the furthest high-probability ($P>0.5$) OC member from \citet{2020A&A...640A...1C}\footnote{It is worth noting that cluster parameters and members in \cite{2020A&A...640A...1C} are derived from \textit{Gaia} DR2, whereas our cluster star samples and stellar properties are obtained from \textit{Gaia} DR3. However, this should not significantly change our sample selection, since we already expand our search to include stars in the extended coronae and outskirts of each cluster.}. 

Next, we compare $r_{prob}$ to the baseline search radius of $50$ parsecs. If $r_{prob} < 50$ parsecs, we utilize the baseline $50$ parsec search radius to encapsulate as many WD+MS binaries as possible. All OCs in our sample adhere to this criteria, with the exception of ASCC 108, Collinder 463, Melotte 20, NGC 6991, Stock 2, UBC 6, UBC 10b, UPK 167 and UPK 294. For these clusters, we instead use a search radius of $r_{prob}$, with the goal of including as many WD+MS binaries as possible while minimizing contamination and computational expense.

\subsubsection{Kinematic Constraints}\label{sec:kinematicclean}

In this study, it is essential that the kinematics of each WD+MS candidate are consistent with those of its suspected host OC. While a complete kinematic analysis is applied to our full catalog in Section \ref{sec:kinematics-candidates}, we initially apply the following set of rough kinematic constraints.

\begin{enumerate}
    \item ${\sigma_{\pi}}/{\pi} < 0.5$: To ensure at least moderately reliable distances to each candidate, we only select stars with relative parallax errors $< 50\%$.
    \item $(\pi, \mu_{\alpha}, \mu_{\delta}) < 1.3 \times P(>0.5)$ members: Similar to Section \ref{sec:searchradius}, we use the high-probability ($P>0.5$) cluster members in \cite{2020A&A...640A...1C} to define a coarse kinematic range for our stellar samples. Using each OC’s proper motion and parallax distributions, we only select stars that have kinematics within a range that extends 30\% beyond the minima and maxima high-probability cluster member values. 
\end{enumerate}

We note that when searching for unresolved binaries in \textit{Gaia}, it is often common to use constraints on the Renormalized Unit Weight Error \citep[\texttt{ruwe}; e.g.][]{2020MNRAS.496.1922B, 2024arXiv240414127C}. A \texttt{ruwe} of 1.0 is expected for single star systems, where \texttt{ruwe} values $> 1.4$ indicate a single star model is a poor fit to the astrometric observations from \textit{Gaia} (and hence, an unresolved multiple star system is possible). However, post-CE envelope binaries are believed to have extremely short orbital periods on the order of a few hours to a few days \citep{2011A&A...536A..43N}. Small orbital separations would thus likely not be detected in the astrometric solution. As highlighted in \cite{2024arXiv240312146E}, \textit{Gaia} light curve data would be most powerful to identify short-period post-CE binaries, however, the sample of systems with light curve data is currently limited (but will be heavily expanded upon with the fourth data release of \textit{Gaia}). While we do not consider \texttt{ruwe} when selecting stars for further analysis, we do provide \texttt{ruwe} values for each of the WD+MS candidates in our final catalog (Section \ref{sec:finalcands}). 

\subsubsection{Photometric Constraints}\label{sec:photclean}

Since we are photometrically identifying WD+MS candidates, we apply the following set of constraints.

\begin{enumerate}
    \item \textit{Gaia} $BP-RP$ $\leq 2.6$: WD+MS binaries typically lie bluewards of the zero-age MS (ZAMS). This is especially evident in \cite{2010MNRAS.402..620R}, who present a catalog of 1602 spectroscopically confirmed WD+MS systems with SDSS DR6 data. When cross-matching this catalog with \textit{Gaia} DR3, we find that 1468 systems have $BP-RP$ color measurements, $> 99.5\%$ of which have $BP-RP$ $< 2.6$.
    \item \textit{Gaia} $m_{G} \leq 19.5$ \& $M_{G} \geq 6.0$: Depending on the distance and relative flux contributions of each binary component, WD+MS systems can exhibit a wide array of magnitudes. While the limiting \textit{Gaia} DR3 apparent magnitude is $m_{G} \sim 21$ \citep{2016A&A...595A...1G}, follow-up spectroscopy of WD+MS binaries with $m_{G} \geq 19.5$ mag is difficult. Furthermore, we find that $> 98.5\%$ of \cite{2010MNRAS.402..620R} WD+MS systems have absolute \textit{Gaia} DR3 magnitudes fainter than $M_{G}=6$.
\end{enumerate}

To define each OC's stellar sample, we query \textit{Gaia} DR3 using \texttt{pyia} \citep{adrian_price_whelan_2018_1228136} within search radii from Section \ref{sec:searchradius}, selecting only stars that pass the kinematic and photometric filters from Section~\ref{sec:kinematicclean} and \ref{sec:photclean}.  

\subsubsection{Cross-matching}\label{sec:crossmatch}

For every OC in Section \ref{sec:openclusters}, we have a sample of \textit{Gaia} DR3 stars from Section \ref{sec:starsample} that are kinematically similar to the host cluster and in the broad region of the \textit{Gaia} color-magnitude diagram (CMD) where WD+MS binaries are expected to appear. However, other classes of objects can also appear in this region of a \textit{Gaia} CMD (see e.g.\ Section~\ref{sec:contaminants}). Still, WD+MS systems exhibit specific ultraviolet, optical and infrared colors due to the presence of a hot WD and a cool M-dwarf companion \citep[e.g.][]{2010MNRAS.402..620R}. We therefore impose the added requirement that sufficient multi-band photometry is available to carry out a detailed assessment of their possible binary nature. To span a range of wavelengths, we cross-match each of our \textit{Gaia} DR3 OC stellar samples with data from Pan-STARRS1 DR1 \citep{2012ApJ...750...99T, 2016arXiv161205560C} and the Two Micron All Sky Survey \citep[2MASS;][]{2006AJ....131.1163S} using \texttt{TOPCAT} \citep{2005ASPC..347...29T}. We adopt the default cross-matching parameters in \texttt{TOPCAT}, using a cross-match radius of 1.0 arcsecond.  We note that two clusters (UBC 577 and UPK 303) had zero cross-matches, so they are discarded from our sample. Finally, we only select stars that have greater than three-sigma detections ($\sigma_{mag} < 0.36$) in all photometric bands to ensure the data is high quality.

While the inclusion of shorter-wavelength photometric bands (i.e. $u$-magnitudes) would allow for detection of hotter WDs (and would better separate WD+MS binaries from MS stars in color-color spaces), existing $u$-band data for OCs is limited. For instance, SDSS DR16 \citep{2020ApJS..249....3A}---which contains $u$-band photometry---only offers sporadic coverage of the Galactic plane (where most OCs are located). Specifically, less than a few dozen of the 301 clusters that meet our criteria outlied in Section \ref{sec:openclusters} retain $> 50\%$ of their stars when cross-matching their stellar samples with SDSS DR16. Other ultraviolet/$u$-band photometric surveys like the Galaxy Evolution Explorer \citep[GALEX;][]{2005ApJ...619L...1M} and the Ultraviolet Near Infrared Optical Northern Survey \citep[UNIONS;][]{2017ApJ...848..128I} also primarily observe the Galactic halo. Future surveys like the Ultraviolet Explorer \citep[UVEX;][]{2021arXiv211115608K} will provide all sky ultraviolet coverage, allowing for incorporation of bluer wavelengths.

Hence, our final stellar samples of 299 OCs contain $G, BP, RP$ photometry from \textit{Gaia} DR3, $g, r, i, z$ photometry from Pan-STARRS1 and $J, H, K_{s}$ photometry from 2MASS. Potential issues when cross-matching between multiple surveys---with different spatial resolutions---within the moderately crowded regions of OCs is addressed in Section~\ref{sec:additionalvetting}.

\begin{figure*}
    \centering
    \includegraphics[width=0.82\textwidth]{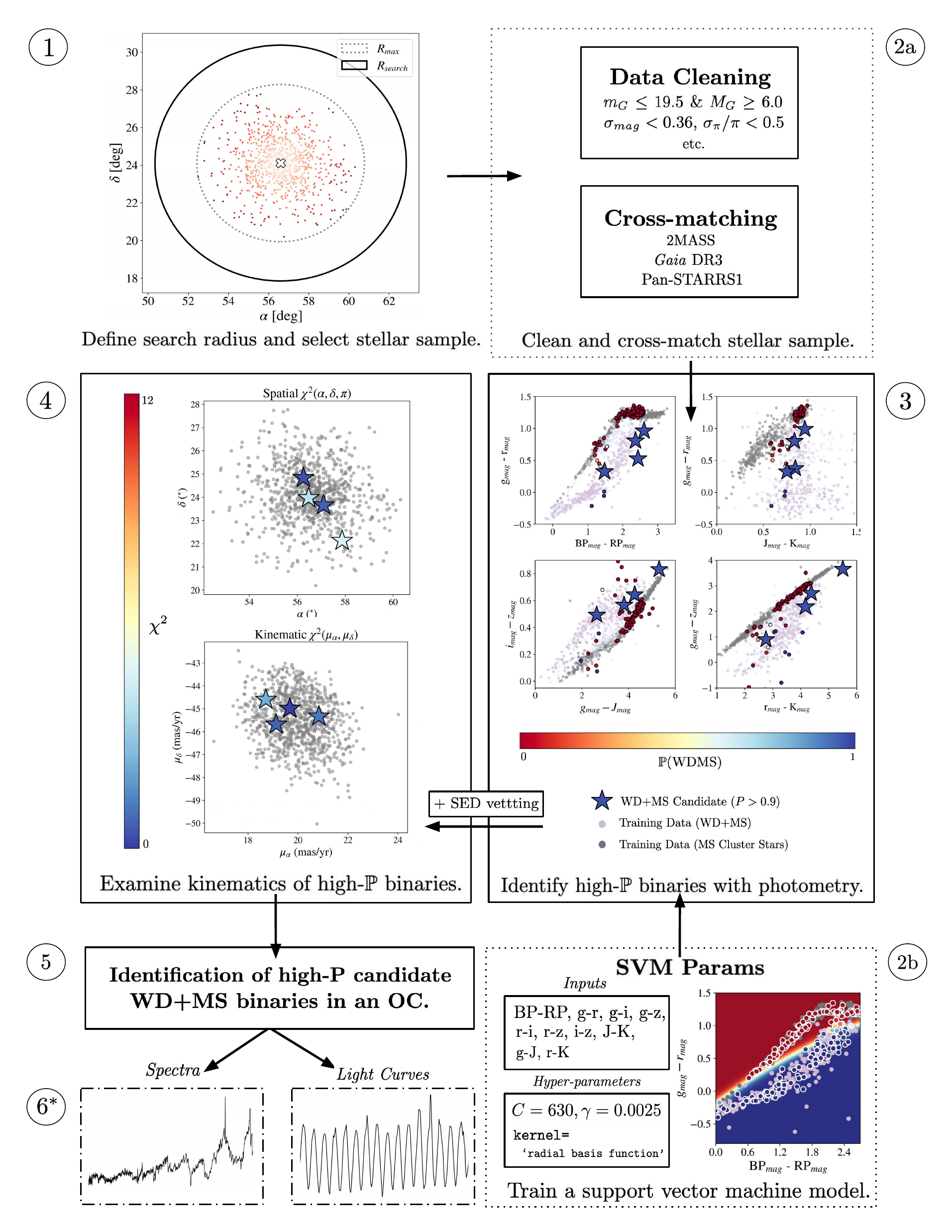}
    \caption{A schematic summarizing our methodology to identify high-probability WD+MS binary candidates in clusters, using the Pleiades as an example. (1) We define an initial \textit{Gaia} DR3 stellar sample utilizing high-probability cluster members from \cite{2020A&A...640A...1C}. (2a) We pre-process and cross-match the  stellar sample with Pan-STARRS1 and 2MASS. (2b) We train a support vector machine classification model to identify stars that are photometrically similar to a known sample of WD+MS systems in SDSS. (3) We run the stellar sample from 2a through our SVM in 2b. The SVM computes the probability that each star in the sample is a WD+MS binary from high-dimensional input photometry. (4) For each candidate identified with $P>0.9$, we first vet its SED and then examine the spatial locations and kinematics relative to a sample of known cluster members. (5) High-probability candidate WD+MS binaries are identified. (6*) We perform follow-up spectroscopy and/or analyze existing light curve data for an example subset of high-probability systems.}
    \label{fig:schematic}
\end{figure*}

\section{WD+MS Candidate Identification} \label{sec:initcand}

As mentioned in Section \ref{sec:crossmatch}, WD+MS binaries exhibit unique photometry and occupy unique regions of CMDs and color-color spaces. Previous studies that identified WD+MS binaries in the field \citep[e.g.][]{2004A&A...419.1057W, 2010MNRAS.402..620R, 2014A&A...570A.107R, 2021MNRAS.506.5201R, 2023MNRAS.tmp.3454N} combined multi-wavelength observations in a variety of CMDs and color-color spaces, highlighting that these binaries are generally photometrically distinct from the majority of MS stars. Here, we present a supervised machine learning method to identify high-probability WD+MS binary candidates based on a variety of input colors. We adopt a machine learning as opposed to manual approach because, in providing probabilities, it can better handle edge cases and/or stars with individual photometric bands that are unphysical (e.g. due to catalog errors). The machine learning method, training data, input features, model definition/evaluation, application to stellar samples and additional vetting of candidates are described below. The overall candidate identification procedure is summarized in Figure \ref{fig:schematic}.

\subsection{Support Vector Machines}

Classification in astronomy often requires combining data from a myriad of surveys and instruments, resulting in high-dimensional datasets. Supervised machine learning approaches are beneficial for classification tasks, since they use known labelled (`training') data to predict the class of objects in an unknown sample. A popular supervised machine learning approach is a support vector machine (SVM). Developed by \cite{cortes1995support}, a SVM creates a $N$-dimensional hyperplane to separate classes of data with $N$-dimensional input features. The SVM hyperplane is defined by a set of `support vectors', which are the points that maximally separate the different classes of input data from one another. Ultimately, the hyperplane represents a decision boundary, classifying data depending on where the points lie relative to the boundary.

SVMs are advantageous when working with non-linear data, as they allow for the definition of non-linear hyperplanes through different kernel choices\footnote{\href{https://scikit-learn.org/stable/modules/svm.html}{https://scikit-learn.org/stable/modules/svm.html}}. Due to the regular presence of non-linearity in astronomical data, SVM approaches have been used to classify a variety of astrophysical phenomena, such as stars, galaxies, active galactic nuclei, quasars and gravitational lenses \citep[e.g.][]{2013A&A...557A..16M, 2017MNRAS.471.3378H, 2022A&A...659A.144W}. For a brief review of SVM applications in astronomy, see \cite{2014ASPC..485..239Z}. We note that while SVMs are widely used in classification, there are a variety of other robust supervised machine learning techniques. For instance, in the binary classification landscape, \cite{Neugent2020} used the k-nearest neighbours method to identify red supergiants with probable O- or B-type binary companions.

\subsection{Training Data} \label{sec:training}
Like other supervised machine learning methods, a SVM requires a set of training data that contains known labelled objects. Since we wish to identify WD+MS binaries, our training data includes two samples: (i) a set of stars that are likely non-WD+MS cluster members and (ii) a population of known WD+MS binary systems. 

\subsubsection{Cluster Star Training Sample}

\begin{figure*}
    \centering
    \includegraphics[width=1\textwidth]{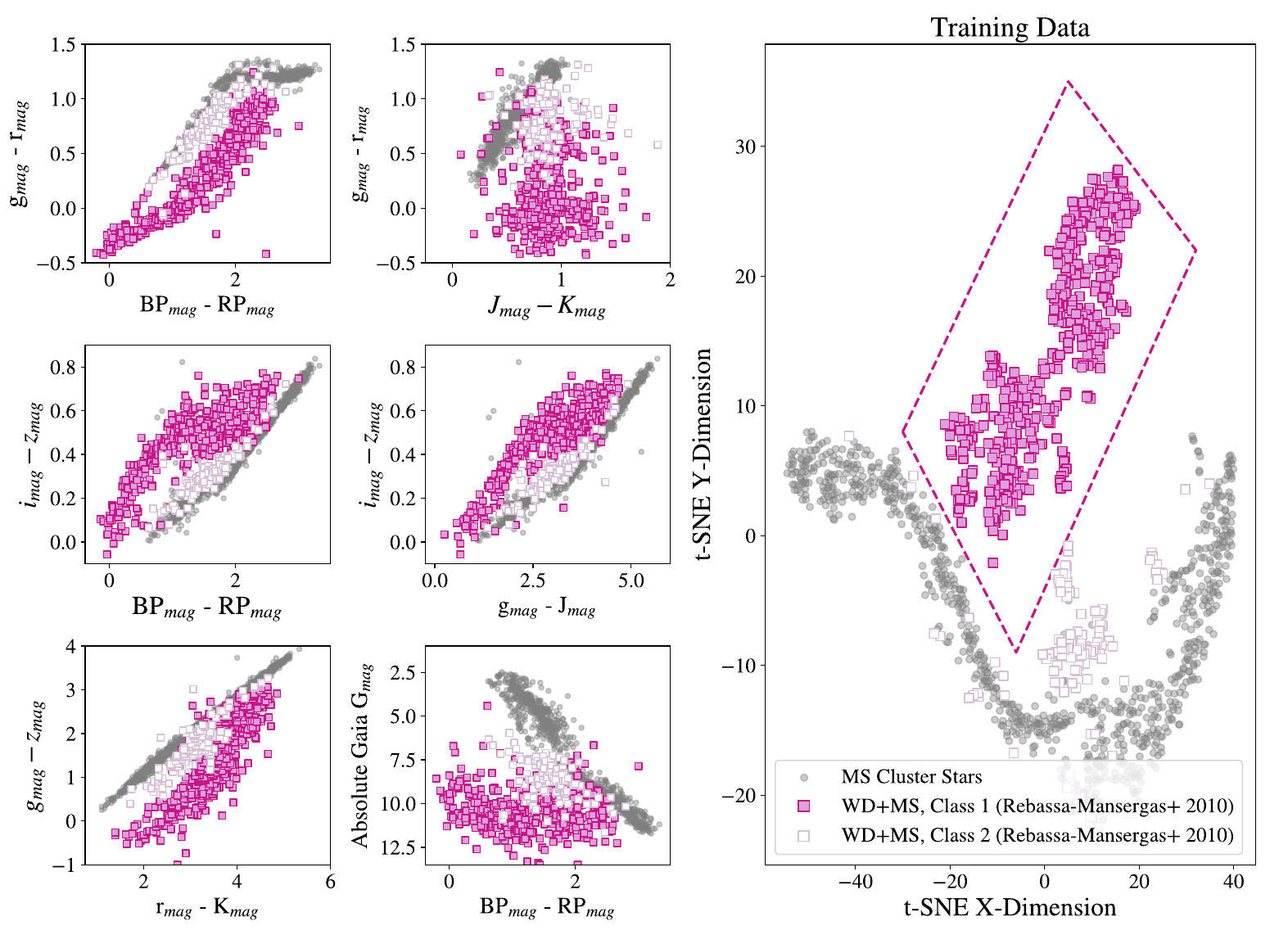}
    \caption{Properties of the training data used to identify WD+MS binaries. MS cluster members from five OCs (Alessi 10, ASCC 41, NGC 2251, NGC 2632 and NGC 7086) are indicated as gray circles. SDSS WD+MS binaries from \cite{2010MNRAS.402..620R} are marked as squares, where the sample is split into two classes. Class 1 (dark pink outlined) represents WD+MS binaries that are distinct from the MS cluster members in a t-SNE projection, five example color-color spaces and a \textit{Gaia} CMD, while Class 2 (light purple) represent a subset of binaries that are more likely to share properties with MS cluster members. To maximize separation between the MS stars and WD+MS binaries and thus minimize potential contamination, we only use Class 1 WD+MS binaries when training our SVM.}
    \label{fig:trainingdata}
\end{figure*}

We first require a set of stars that can be used to describe the expected properties of \emph{non}-WD+MS stars. Since we are searching within OCs, we select a comparison sample of stars from known OC members. We select stars from five OCs (Alessi 10, ASCC 41, NGC 2251, NGC 2632 and NGC 7086) that span a variety of ages ($\sim 90-700$Myr) and extinctions ($A_{V}\sim0-2$) representative of our OC sample described in Section \ref{sec:openclusters}. For inclusion in the training sample, we require that a star was identified as having a membership probability of $P>0.7$ in \cite{2020A&A...640A...1C} and that, after cross-matching, it has greater than three-$\sigma$ detections in all photometric bands. In addition, we exclude a set of bright ($g<$14 mag) stars located in NGC 2632 where data-quality issues due to saturation in Pan-STARRS1 were identified.
This leaves  875 cluster members in our training sample: 42 cluster stars in Alessi 10, 39 stars in ASCC 41, 57 stars in NGC 2251, 379 stars in NGC 2632 and 358 stars in NGC 7086.

While we do not remove potential WD+MS cluster binaries from this sample, after running our full classification method, we identify only one cluster member with a SVM probability of $>50\%$ and zero cluster members with $90\%$ probabilities of being WD+MS stars among this sample.  As this source represents only $0.2\%$ of the total sample and a SVM works by defining a plane between two populations, we expect the effect of potential binaries in our cluster member sample to be minimal.

In addition to our cluster member sample, we also include a set of spectral type B8V to M8V \cite{1998PASP..110..863P} model stars. Defining a boundary where the bluest possible non-WD+MS binaries could exist is important for yielding a SVM hyperplane that properly separates WD+MS binaries from bluer MS stars. While we already include low-reddening clusters in our training sample, incorporating zero-reddening ($A_{V}=0$) \cite{1998PASP..110..863P}  stellar models ensures that we span the bluest possible range for all MS spectral types that could be picked up in our stellar samples. Synthetic Pan-STARRS1, 2MASS and \textit{Gaia} photometry for 29 model stars is obtained after scaling the fluxes for the models to match the \textit{Gaia} absolute magnitudes in the ``Modern Mean Dwarf Stellar Color and Effective Temperature Sequence'' distributed by E. Mamajek\footnote{\url{https://www.pas.rochester.edu/~emamajek/EEM_dwarf_UBVIJHK_colors_Teff.txt}} \citep{Pecaut2013}. 

\subsubsection{WD+MS Training Sample}\label{sec:wdmstraining}

Our set of known WD+MS systems is taken from \cite{2010MNRAS.402..620R}, who provide a catalog of 1602 WD+MS binaries from SDSS DR6. After cross-matching this catalog with \textit{Gaia} DR3, Pan-STARRS1 and 2MASS and cleaning the sample using the same constraints as the cluster star samples and training data, we are left with 570 WD+MS binaries.

To explore the global properties of the \cite{2010MNRAS.402..620R} sample (which were identified spectroscopically using a template fitting technique), we use the t-SNE algorithm \citep{JMLR:v9:vandermaaten08a}. T-SNE is an unsupervised machine learning method that reduces the dimensionality of a high-dimensional dataset and has been widely used in astronomy to easily examine and visualize similarities \citep[e.g.][]{2018A&A...619A.125A, 2018MNRAS.473.4612K,2019MNRAS.483.3196C, 2023MNRAS.518.4249G}. Using \texttt{scikit-learn} \citep{scikit-learn}, we run the t-SNE algorithm with an input set of 10 colors from \textit{Gaia} DR3, Pan-STARRS1 and 2MASS with a high perplexity ($p=100$), to identify global rather than local features. The results are plotted in Figure \ref{fig:trainingdata}.

\begin{figure*}
    \centering
    \includegraphics[width=0.9\textwidth]{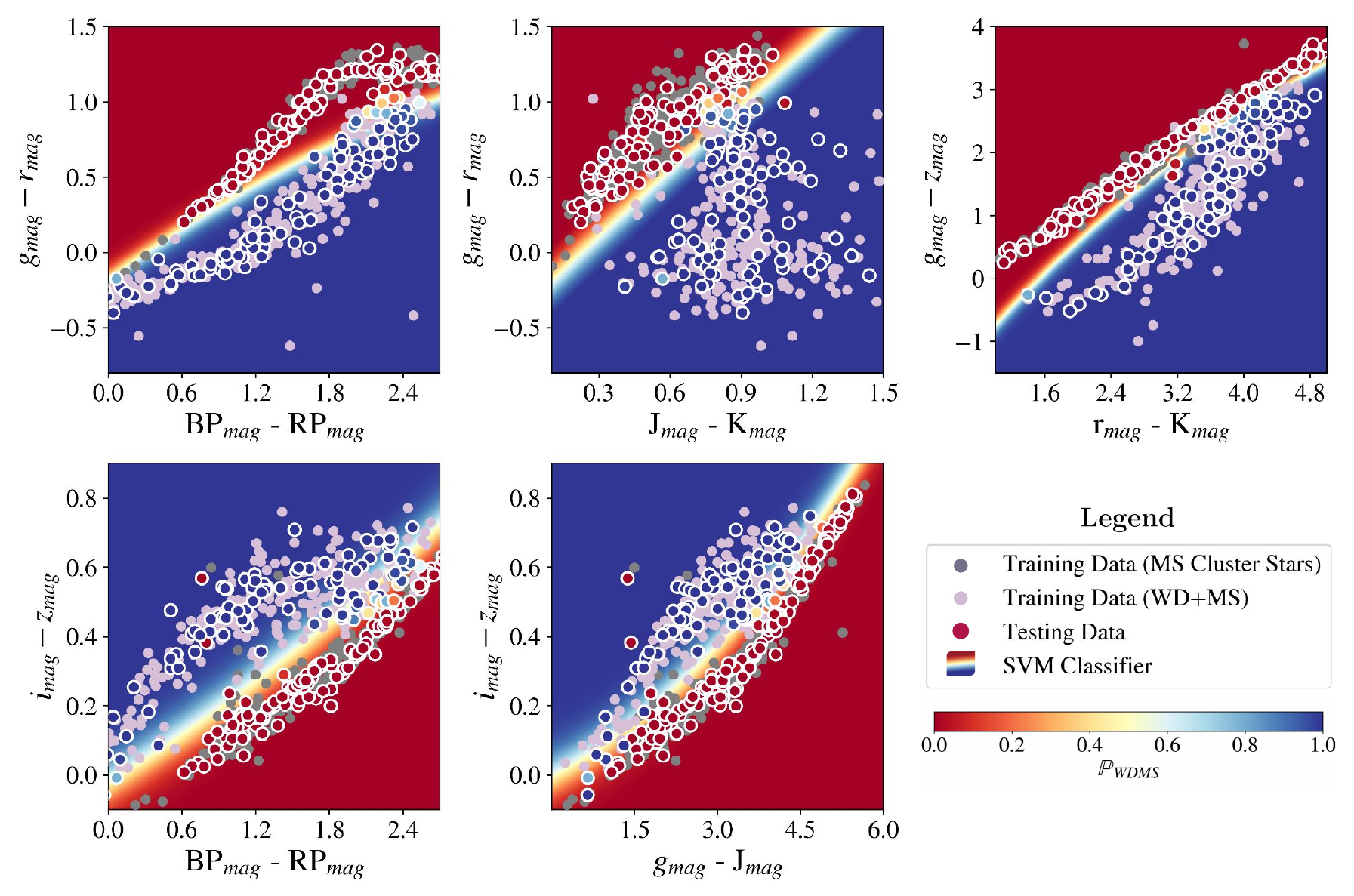}
    \caption{Our support vector machine (SVM) model to identify WD+MS binaries in five different color-color spaces. Our test data is run through the SVM, where each point is colored by its probability of being a WD+MS binary. The training sample containing OC members (gray) and known WD+MS binaries from SDSS \citep[purple,][]{2010MNRAS.402..620R} is highlighted for reference. Overall, the assigned probabilities for the test data are aligned with the SVM classifier (rainbow gradient), where any outliers can likely be explained by higher-dimensional effects (i.e. here, we use 2-D example spaces to visualize the SVM classification, however the SVM is actually a 10-dimensional hyperplane and uses all input colors to compute probabilities).}
    \label{fig:svmexample}
\end{figure*}

From Figure \ref{fig:trainingdata}, we see that most WD+MS binaries are distinct from the cluster star training sample in the two-dimensional t-SNE projection, indicating that the global properties of the population are unique from the MS stars. However, we observe that there is some separation among the WD+MS sample, and we split them into two classes (as defined by the pink box in the right panel of Figure~\ref{fig:trainingdata}). Class 1 consists of 445 WD+MS binaries that are widely separated from the cluster sample in five example CCDs and a \textit{Gaia} CMD. Class 2 consists of 125 WD+MS binaries that lie closer to the cluster stars in all aforementioned spaces. 

The overall redder colors and higher absolute magnitudes of the Class 2 systems may simply be attributed to higher mass MS companions. However, after running a series of tests and multiple iterations of our SVM model, we chose to remove them  from our training sample in order to minimize contamination and maximize separation between the cluster members and WD+MS binaries. We discuss the impact of this choice on our final sample contamination and completeness in Sections \ref{sec:contamination} and \ref{sec:complete} below. Class 1 represents the final training sample of WD+MS binaries in this study.
 
\subsection{SVM Input Features}
As discussed in Section \ref{sec:crossmatch} and shown in Figure \ref{fig:trainingdata}, WD+MS binaries occupy different regimes of CMDs and color-color spaces than MS stars \citep[e.g.][]{2010MNRAS.402..620R, 2021MNRAS.506.5201R, 2022AJ....164..126A, 2023MNRAS.tmp.3454N}. Thus, we train our SVM to identify WD+MS binary candidates using colors constructed from \textit{Gaia} DR3, Pan-STARRS1 and 2MASS magnitudes. Specifically, our set of input features includes 10 colors: $BP-RP$, $g-r$, $g-i$, $g-z$, $r-i$, $r-z$, $i-z$, $J-K$, $g-J$, and $r-K$. While many WD+MS systems are also located in a unique region of various CMDs (appearing bluewards of the ZAMS and redwards of most isolated WDs), we do not input any absolute magnitudes when training our SVM in order to mitigate effects caused by parallax uncertainty. 

\subsection{Model Definition}\label{sec:model-def}

In this study, we implement a SVM model using the \texttt{scikit-learn} \citep{scikit-learn} Python package. We randomly split up our initial training dataset in Section \ref{sec:training} into three sub-groups, using 60\% for training and reserving 20\% for model validation and 20\% for testing. Ultimately, this allows us to validate and test the SVM on new input data to avoid overfitting our model.

To build our SVM, we first use our validation data to tune a set of hyper-parameters to determine the best model parameters for classification. For our SVM, the most important parameters are the regularization ($C$), kernel choice and kernel coefficient ($\gamma$). To determine the optimal model $C$, kernel and $\gamma$, we perform a coarse search over all parameters using \texttt{GridSearchCV} with $k=5$ folds in \texttt{scikit-learn} \citep{scikit-learn}. Using the cross-validation outcomes, we then perform a finer grid search to identify the optimal model parameters. Using our validation data, we find that $C=630.957$ and a radial basis function kernel with $\gamma=0.0025118$ yields the best model.

\begin{figure}
    \centering
    \includegraphics[width=0.9\columnwidth]{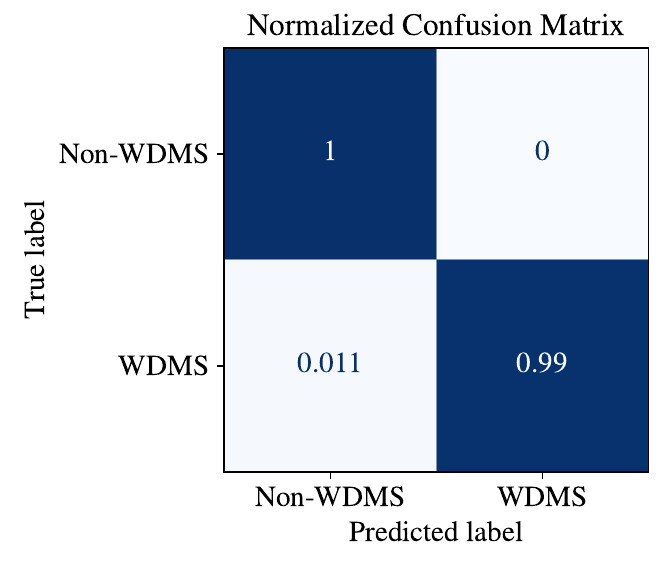}
    \caption{A normalized confusion matrix of our SVM classifier using our test data. Overall, the SVM performs well, with no false positives and only one false negative identified out of 265 systems when utilizing a prediction probability threshold of $P>0.5$.}
    \label{fig:confusion}
\end{figure}

Our SVM contains 10 input features, hence the resulting hyperplane separating the WD+MS population from the likely MS cluster stars is high-dimensional. To visualize the SVM classifier, we highlight some example 2-D parameter spaces in Figure \ref{fig:svmexample}. In each of these color-color spaces, the WD+MS binaries are clearly separated from the MS cluster stars in both the training and testing data. The SVM plane is represented as a gradient, where the color corresponds to the probability that a system is either a WD+MS binary or a MS cluster star. It is worth remembering that the SVM classifies systems by including \textit{all input features}, so a star that appears to be a WD+MS binary but is classified as a MS star likely shares properties consistent with the MS population in higher dimensions (and vice versa). As shown with a confusion matrix in Figure \ref{fig:confusion}, the SVM classifies the 20\% of systems reserved for testing incredibly well. In our testing data, only $\sim 1\%$ of WD+MS binaries are classified as MS stars when adopting a prediction threshold above a probability of $P>0.5$. More importantly, using the same probability threshold, zero cluster members are classified as WD+MS binaries.

\subsection{Model Evaluation}
Although we find that the number of misclassified stars in our testing data is low, we recognize that there are other potential factors that could impact the performance of our model when applied to the sample of cluster stars described in Section~\ref{sec:search}. Below, we perform a series of tests to probe the expected contamination and completeness of our final catalog. Throughout, we emphasize that given only 2 post-CE systems in OCs are currently known, our primary motivation in performing this study is not to be 100\% complete, but rather to identify a set of high-probability WD+MS candidates \emph{with low contamination} that can subsequently be followed-up in more detail.

\newpage
\subsubsection{Catalog Contamination}\label{sec:contamination}

\begin{figure*}
    \centering
    \includegraphics[trim=0.25in 0.8in 0.25in 0.75in , clip, width=0.85\textwidth]{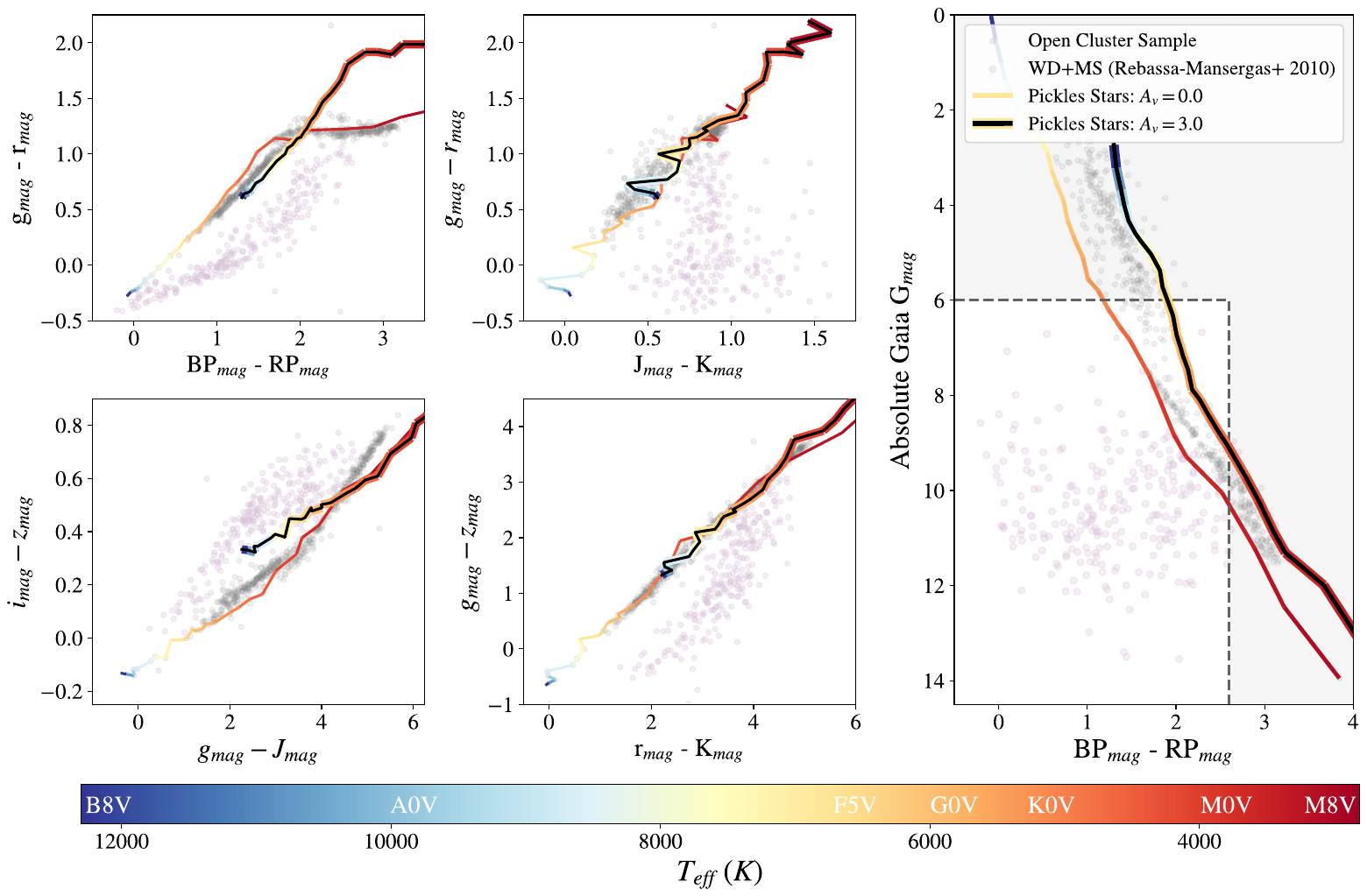}
    \caption{Reddened MS stars of spectral types B8V to M8V relative to the training samples of OC members from \cite{2020A&A...640A...1C} and SDSS WD+MS binaries in \cite{2010MNRAS.402..620R}. The thin line shows the \cite{1998PASP..110..863P} stellar atmosphere models for an extinction of $A_{V}=0$ while the thick black line depicts an extinction of $A_{V}=3$.  Both lines are colored by the effective temperatures of the stars. It is clear that reddening of MS stars moves them along or away from the known WD+MS sample.}
    \label{fig:extinction}
\end{figure*}

\textit{Impact of Extinction:} While our training sample includes members from clusters that span a wide range of extinction values ($0 < A_V < 2$; Section~\ref{sec:training}), we do not explicitly attempt to correct any stars for line of sight extinction. Hence, we examine the impact that this can have on our method of WD+MS identification. In particular, we wish to understand whether a MS star with high reddening could be inadvertently identified as a WD+MS candidate by the SVM model described above. To test this, we use the same \citet{1998PASP..110..863P} stellar atmosphere models as in Section \ref{sec:training} for B8V to M8V MS stars. 
We apply a range of reddening values (0 $<$ A$_{\rm{V}}$ $<$ 3) using a \cite{Cardelli1989} extinction law with R$_{\rm{V}} = 3.1$ and perform synthetic photometry in the Pan-STARRS1 $g, r, i, z$, 2MASS $J, H, K_s$ and the \textit{Gaia} G, G$_{\rm{BP}}$, and G$_{\rm{RP}}$ filters. 

In Figure~\ref{fig:extinction} we show the resulting MS for both A$_{\rm{V}} = 0$ and A$_{\rm{V}} = 3$ in the \textit{Gaia} CMD and four representative color-color diagrams. The training sample of cluster and WD+MS stars described in Section~\ref{sec:training} are also shown for context. In many of the color-color spaces, increasing the reddening of a MS star moves its location either \emph{along} the MS or \emph{away} from the location of the known WD+MS sample. To quantitatively test this, we run the reddened Pickles models through our SVM model. None of these stars are identified as a probable WD+MS system, with  P$<$0.20 in all cases.

\newpage

\textit{Impact of WD+MS Training Sample:} As described in Section~\ref{sec:wdmstraining}, after running t-SNE we separated the WD+MS sample from \cite{2010MNRAS.402..620R} into two subsets and used only the Class 1 objects in our baseline SVM model. However in addition, we trained a separate SVM model using the same procedures described in Section~\ref{sec:model-def} in which we utilized both the Class 1 and Class 2 WD+MS binaries. While the model still performed well, we found that the number of misclassified MS cluster members at intermediate probabilities increased. Since the overall number of MS stars is orders of magnitudes larger than the number of expected WD+MS binaries in clusters, contamination of even a few percent could cause falsely identified systems to dominate over real WD+MS systems. Hence, we opt for our baseline SVM model trained with Class 1 binaries from \cite{2010MNRAS.402..620R}.

\begin{figure*}
    \centering  \includegraphics[width=\textwidth]{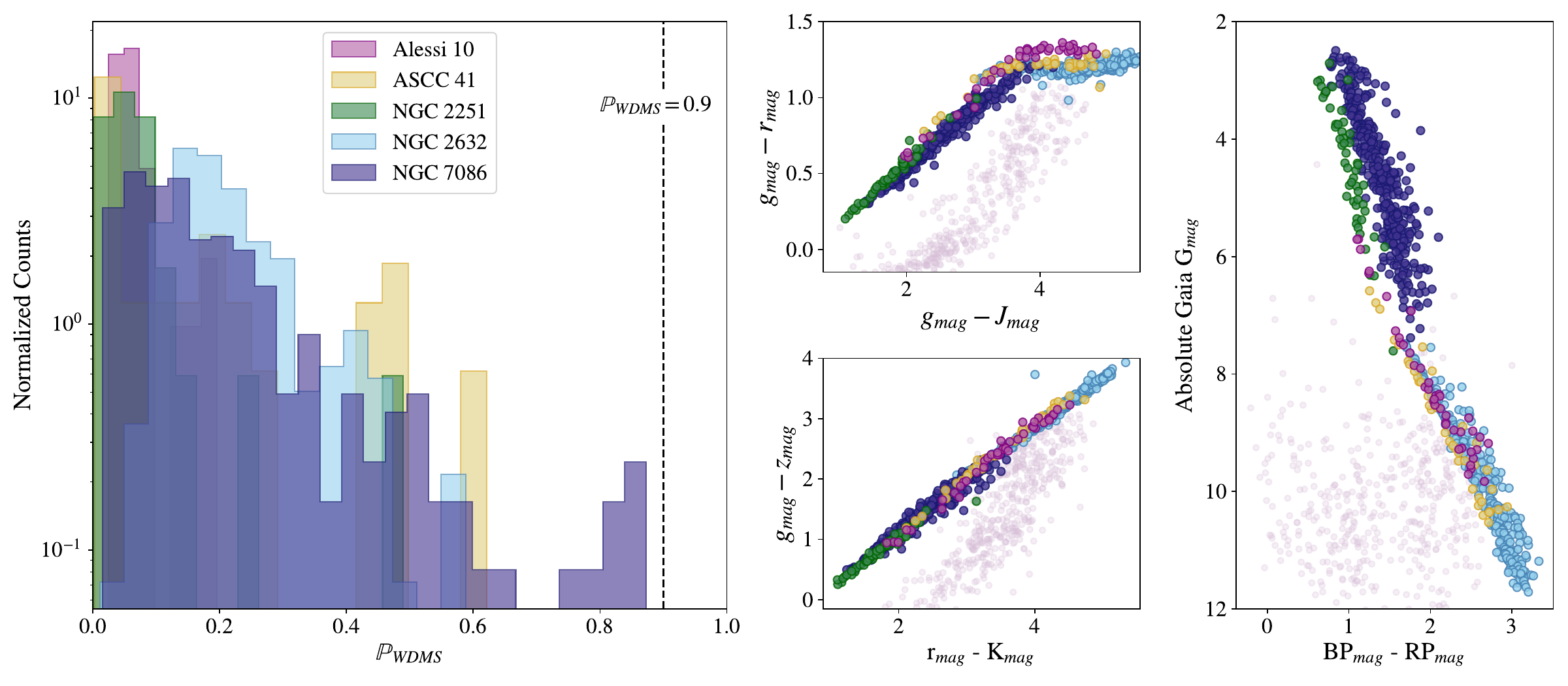}
    \caption{Leave-one-out cross validation (LOO-CV) results for our training sample. To investigate the performance of our SVM on `new' clusters, we remove one cluster from our training sample and use it as test data with our SVM (each left out cluster is labelled). Overall, the SVM classifies the cluster members as low-probability WD+MS systems, where only a few cluster members are classified as high-probability WD+MS binaries. Specifically, zero sources are classified with probabilities above $P>0.9$. Hence, we opt to use this as a metric for defining a high-probability sample of systems in this study. The OC with the highest cluster member contamination is NGC 7086 (dark purple), which is due to its broad colors extending near the known WD+MS binaries (light purple) in color-color spaces.}
    \label{fig:loocv}
\end{figure*}

\textit{Impact of Specific Cluster Properties:} As described in \ref{sec:training}, our training sample is composed of both cluster members from five OCs and model stars spanning a variety of spectral types. While this sample does encompass a wide range of extinctions and colors, it is possible that some OCs investigated in this study have different overall properties than the stars used in our training data. To mimic the process of classifying an unseen cluster with our SVM, we perform a leave-one-out cross validation (LOO-CV) test on our training data. Using the \texttt{LeaveOneGroupOut} function from \texttt{scikit-learn} \citep{scikit-learn}, we remove one cluster from our training data and run it through our SVM model, repeating this for each of the five OCs.

Figure \ref{fig:loocv} highlights the results of the LOO-CV, where each labelled cluster is the one that was left out of training and used as the test data with the SVM. Overall, the computed WD+MS probabilities of these cluster members are relatively low, where some clusters (i.e. Alessi 10 and NGC 2251) have little contamination even at medium probabilities (i.e. 0\% of members are classified as WD+MS binaries at $P>0.5$). 
Other clusters like ASCC 41, NGC 2632 and NGC 7086 yield larger misclassification rates (2.5\%, 0.83\% and 4.8\% at $P>0.5$, respectively). However, this is not unexpected: in various parameter spaces these clusters had colors and magnitudes that were ``closest'' to the sample of WD+MS stars (see Figure~\ref{fig:loocv}). Removing them from the training sample is therefore likely to shift the location of the hyperplane which the SVM uses to separate the samples (as in Figure~\ref{fig:svmexample}).

We emphasize that we expect our final model to have lower contamination than was found based on these LOO-CV tests as: (i) we specifically chose these five clusters to span the range of observed behaviour for the full sample and (ii) in the case of NGC 7086, two-thirds of the stars misclassified with $P>$0.6 are bright MS stars with $M_{\rm{G}} < 6$ mag, which would be excluded from our sample based on the cuts described in Section~\ref{sec:photclean}.  However, even with these caveats, we note that no cluster source in these LOO-CV tests was classified with a WD+MS probability larger than $P>0.9$, indicating that contamination from MS stars is likely minimal above this probability threshold. With this as motivation, we conservatively present only \textit{high-probability} candidates identified with $P>0.9$ in our catalog (see Section \ref{sec:application}).

\subsubsection{Catalog Completeness}\label{sec:complete}

Any photometric method to identify WD+MS stars will not be complete---especially when UV photometry is not available---because for certain combinations of binary masses and ages, the MS companion will completely dominate the SED. In addition, as described above, we purposely design our SVM model to minimize contamination from non-WD+MS sources by removing a subset of the SDSS WD+MS systems (Class 2) from \cite{2010MNRAS.402..620R} that lie close to the ZAMS in both a \textit{Gaia} CMD and multi-band color color spaces (Figure \ref{fig:trainingdata}). However, as a consequence, our method may be less sensitive to WD+MS binaries with more luminous (massive) MS stars.

\begin{figure}
    \centering  \includegraphics[width=\columnwidth]{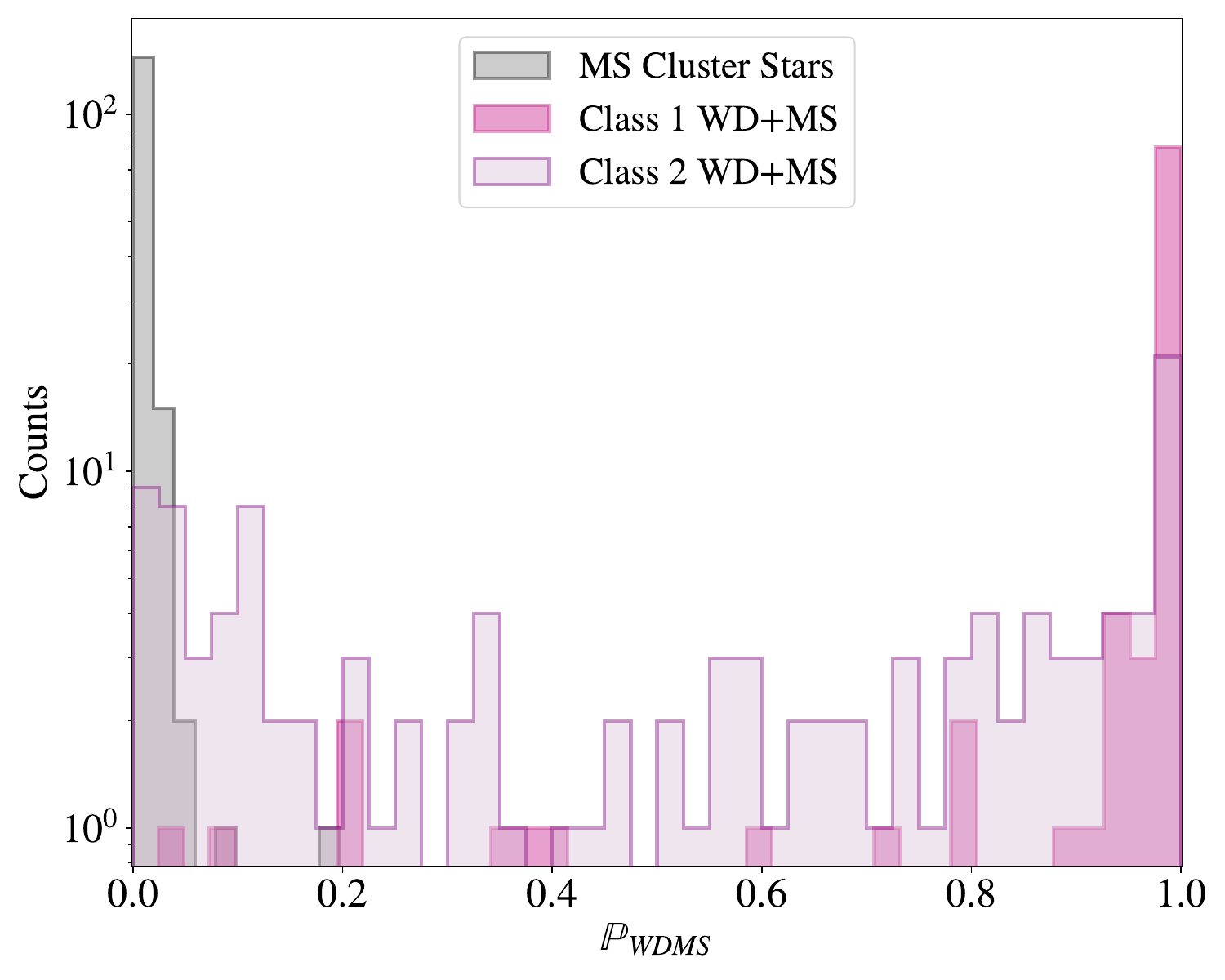}
    \caption{SVM probabilities for sub-samples of systems (not) used in our SVM training. As described in Section \ref{sec:wdmstraining}, we use a sample of likely MS cluster members (gray) and WD+MS binaries from \cite{2010MNRAS.402..620R} to train our SVM. To minimize contamination, we only use `Class 1' binaries (dark pink) in our model, which on average have larger separation from the ZAMS in color-color spaces and a \textit{Gaia} CMD than the `Class 2' binaries (light pink). Here, we see that while our model is less sensitive to these types of binaries, $\sim 25\%$ of the `Class 2' systems are still classified with high-probability with our SVM. Hence, these types of binaries could be present in our catalog. }
    \label{fig:wdcomparison}
\end{figure}

To test this, we run the Class 2 WD+MS systems through our SVM described in Section \ref{sec:wdmstraining} and plot the systems relative to our sample of MS cluster stars and Class 1 WD+MS binaries in our testing data in Figure \ref{fig:wdcomparison}. We find that $>25\%$ of the Class 2 WD+MS binaries are still identified with probabilities of $P>0.9$. Hence, while we do sacrifice some completeness towards systems with high-mass companions in our training sample, our model should still be sensitive to some of these binaries.

In addition, we assume that the colors of WD+MS field binaries in \cite{2010MNRAS.402..620R} will be similar to those of WD+MS systems in clusters. However, for particularly young clusters, it is possible that low mass M-dwarfs will still be in the pre-MS phase, possibly reducing our sensitivity to those systems. However, as seen in \cite{2012AIPC.1480...22P}, almost all $0.1 \mathrm{M_\odot}-0.2 \mathrm{M_{\odot}}$ dwarfs evolve onto the ZAMS within $\sim 50-100$Myr. Hence, while younger OCs in our sample may be slightly less complete towards systems with very low mass companions, our method should be able to identify systems with low mass dwarfs in OCs with ages $\gtrsim 100$Myr.

\newpage 
\subsection{Application to Cluster Data}\label{sec:application}
We run our SVM classification model on 52,556 stars associated with 299 OCs that were selected as described in Section \ref{sec:search}. Based on the results of our LOO-CV, we deem any source identified with a SVM probability larger than $P>0.9$ to be a high-probability candidate WD+MS binary and save it for further inspection. In 234 clusters ($\sim$78\%), our model returns \emph{zero} high-probability candidates. In the remaining clusters, there are typically $1-4$ systems identified with $P>0.9$, with one notable exception: when run on stars in the region surrounding NGC\,1977, our model returned 75 candidates with a high-probability of being a WD+MS system. After visual inspection, we found that this cluster is located in a region of very high extinction, and that significantly more scatter was present in the Pan-STARRS1 photometry than was found in our training sample. We therefore consider the results from this cluster unreliable and exclude it from our sample. After removing NGC\,1977, we are left with 118 candidates in 64 clusters that are classified with high-probability by the SVM. Example plots of SVM results for clusters which both do and do not have any high-probability WD+MS candidates are shown in Appendix \ref{ap:example-search}.

\subsection{Additional Vetting of  Candidates}\label{sec:additionalvetting}

While we already removed sources with large magnitude errors ($\sigma_m > 0.36$) from our stellar samples, additional systematic effects could lead to non-physical SEDs, causing some sources to then be (incorrectly) classified as high-probability WD+MS stars by our SVM model. In particular, it is possible that in the crowded regions of some OCs, the lower angular resolution of 2MASS may lead to a situation where individual 2MASS sources contain the flux from multiple sources which are individually resolved by \textit{Gaia}/Pan-STARRS1. In addition, photometry in the catalogs could be affected by saturation---either of the star itself or due to the image quality being affected by nearby bright objects. We therefore performed a more detailed inspection of the 118 high-probability candidates by examining their SEDs, the Pan-STARRS1/2MASS catalog quality flags, and images of the region surrounding each star.

Upon visual inspection of our SEDs, we noticed that the photometry of many systems behaved as expected: we observed blue ($g$-band) flux upticks of varying degrees (indicative of a hot companion), along with photometry well-fitting late-type MS stars at redder wavelengths (consistent with a MS companion). However, others exhibited abnormalities. In particular, we observed a number of systems that exhibited large flux jumps between adjacent bands, especially when transitioning from optical to infrared wavelengths--potentially indicative of the cross match issue described above.

\begin{figure*}
    \centering
    \includegraphics[width=1\textwidth]{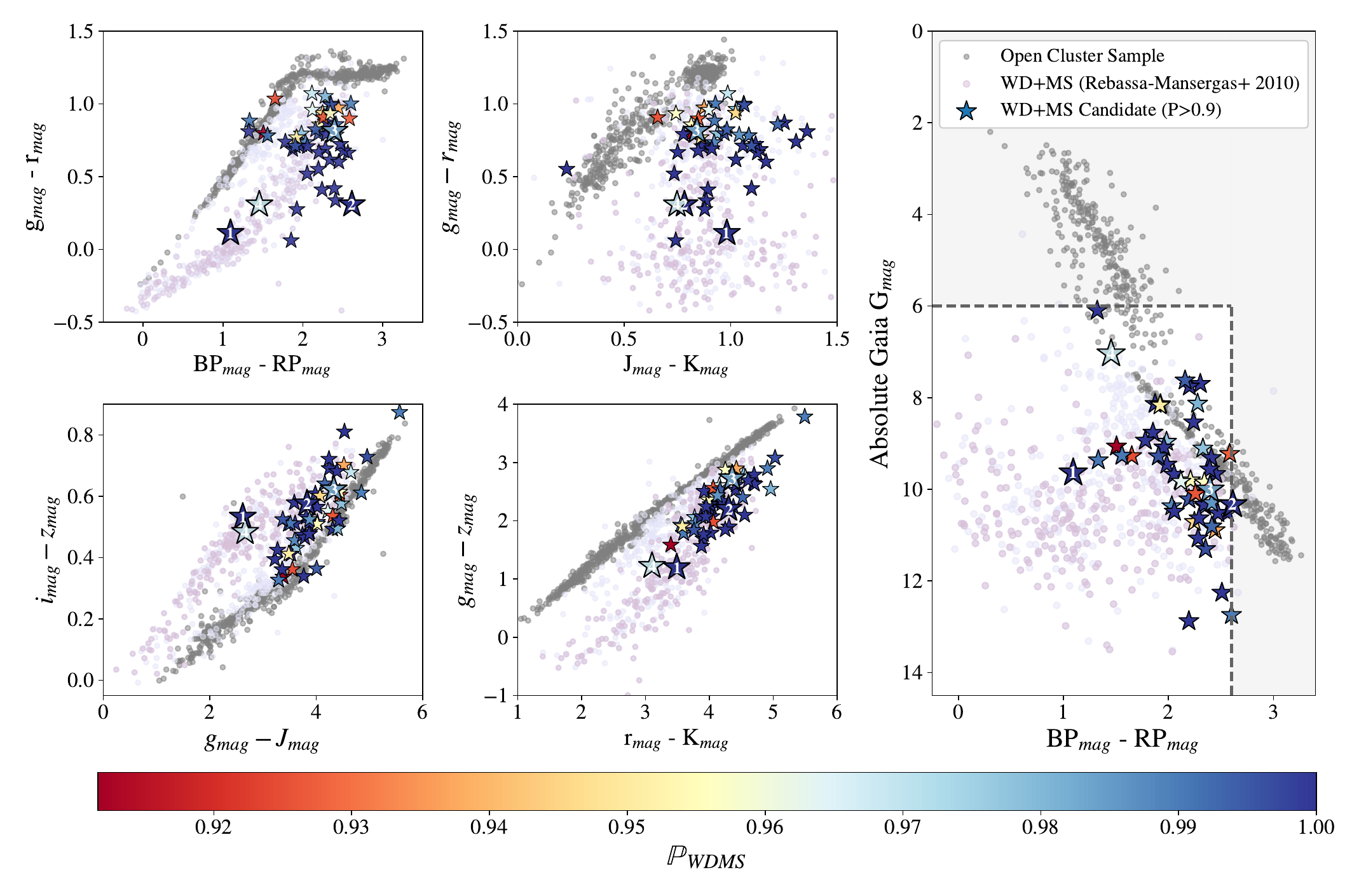}
    \caption{Photometric properties of our catalog of 52 high-probability ($P>0.9$) candidate WD+MS binaries (stars). Each system is colored by its SVM probability. Four high-probability candidates with spectroscopy and/or a light curve discussed in Section \ref{sec:verification} in this paper are numbered: (1) Alessi12-c1, (2) Pleiades-c1, (3) Pleiades-c2, (4) Pleiades-c3. The MS cluster training data (gray) and the \cite{2010MNRAS.402..620R} SDSS WD+MS sample (purple) are also shown for reference. Note that the WD+MS training data used in our SVM (only Class 1 WD+MS binaries; see Section \ref{sec:wdmstraining}) is indicated in plum, whereas the full \cite{2010MNRAS.402..620R} sample (including both Class 1 and Class 2 WD+MS systems) is indicated in light purple.  While the systems in our catalog generally overlap with the SDSS binaries, our candidates lie closer on average to the ZAMS in a \textit{Gaia} CMD and some example color-color spaces.}
    \label{fig:highprobcandidates}
\end{figure*}

We used multiple methods to identify which subset of our candidates may be affected. First, using the same set of B8V to M8V Pickles models (with $0 < A_{V} < 3$) described in Section \ref{sec:contamination} as a baseline, we selected stars that had non-physical $i-z$, $z-y$, $y-J$ or $J-K$ colors (where we define ``non-physical'' as colors beyond the distribution of the Pickles MS models). Second, we inspected the Pan-STARRS1 color image around the system using the Aladin Lite interactive sky atlas \citep{2000A&AS..143...33B}, over-plotting the locations of both the \textit{Gaia} and 2MASS measurements relative to the star of interest. We identified multiple cases where either (i) there were multiple \textit{Gaia} sources of similar brightness close to a single 2MASS source or (ii) where there was a clear red source in very close proximity to a blue star in the Pan-STARRS1 image. In total, we identified 55 systems where we are not confident that the SED fed into our SVM model is representative of a single stellar system. We removed these from our candidate sample.

In addition, when investigating Pan-STARRS1 photometric quality flags, we found that nine systems exhibited large magnitude fluctuations (standard deviations $>0.3$ mag) between multiple Pan-STARRS1 measurements in either the $g$, $r$ or $z$-bands. Upon inspecting the region around each star, we found that many of these sources were located near very bright stars that are oversaturated in the Pan-STARRS1 images. Hence, we also removed these systems from our catalog. Finally, we carefully inspected each SED for sources with Pan-STARRS1 $g$-band magnitudes brighter than $g=14$ mag, as sources above these magnitudes can be impacted by saturation. We removed two additional objects from our catalog due to potential saturation effects.

In summary, after manually inspecting the photometry and on-sky positions of all of our candidates, we are left with \textit{52 high-probability candidate WD+MS binaries in 38 OCs}. SEDs for all of our vetted systems are presented in Appendix \ref{sec:seds}.

\newpage
\section{Candidate Characteristics} \label{sec:photocand}

Here we describe the photometric, spatial and kinematic properties of the 52 high-probability ($P>0.9$) vetted WD+MS candidates identified in Section \ref{sec:initcand}.

\subsection{Photometric Properties}\label{sec:photometry}

Figure \ref{fig:highprobcandidates} highlights our sample of 52 high-probability binaries relative to our training samples of SDSS WD+MS systems \citep{2010MNRAS.402..620R} and MS cluster members (a version of this plot with photometric errors included is shown in Figure~\ref{fig:photerrors} of  Appendix~\ref{ap:photo-errs}). Our candidates broadly overlap with the SDSS sample of WD+MS stars in these parameter spaces, as expected based on our SVM method described above. However, two points warrant further note.

\clearpage

First, we find that 14 ($\sim27\%$) of our high-probability candidates either overlap with, or are found slightly redward of the MS cluster stars in our training sample in the \textit{Gaia} CMD. We emphasize that we provided only color information, and not absolute magnitudes, to the SVM during the model training/candidate identification. These systems may represent stars with (i) uncertain distances (thus impacting their position on the vertical axis of Figure~\ref{fig:highprobcandidates}), (ii) locations in regions of higher reddening or (iii) slightly more evolved MS companions. We note that the spectroscopic sample of SDSS WD+MS binaries from \cite{2010MNRAS.402..620R} does also contain a few stars in this region.

Second, we highlight that \emph{on average}, our sample of candidate WD+MS systems is closer to the ZAMS in the \textit{Gaia} CMD than the SDSS WD+MS sample. In particular, $\sim40\%$ of our high-probability candidates are located within $\pm 0.25$ mag (in $BP-RP$) of the ZAMS, as compared to $<20\%$ of the SDSS WD+MS sample from \cite{2010MNRAS.402..620R}. A similar trend is observed in our sample color-color spaces, where our candidates tend to overlap with the redder portion of the distribution of SDSS WD+MS binaries. As the location of a given WD+MS system in these parameter spaces depends on the flux ratio and temperatures of the WD and MS stars, this effect is likely caused by a combination of the true prevalence of different types of systems as well as the relative sensitivity of our method (which relies on photometry) versus that of \cite{2010MNRAS.402..620R} (which relies on spectroscopy) for detecting different combinations of WD+MS systems.

Indeed, \cite{2021MNRAS.506.5201R} observe a similar effect when searching for WD+MS stars in the Solar Neighbourhood using \textit{Gaia}. They find that their \textit{Gaia}-identified systems generally contain cooler WDs when compared to the SDSS sample of \cite{2010MNRAS.402..620R}. They hypothesize this is an observational effect where \textit{Gaia} is better able to detect binaries where flux contributions from cool WDs are comparable to flux contributions from cool M-dwarf companions in comparison to the spectroscopy template matching method of \cite{2010MNRAS.402..620R}. Cooler WD primaries are also found in \textit{Gaia} WD+MS binaries identified by \cite{2021MNRAS.504.2420I}. While spectra for each of our candidates would be required to determine effective WD temperatures, binaries with overall redder colors seem to generally be consistent with other WD+MS systems identified in \textit{Gaia}.

\newpage

\subsection{Spatial \& Kinematic Properties}\label{sec:kinematics-candidates}

\begin{figure*}
    \centering
    \includegraphics[trim=0.0in 0.25in 0.in 0.in , clip,width=0.95\textwidth]{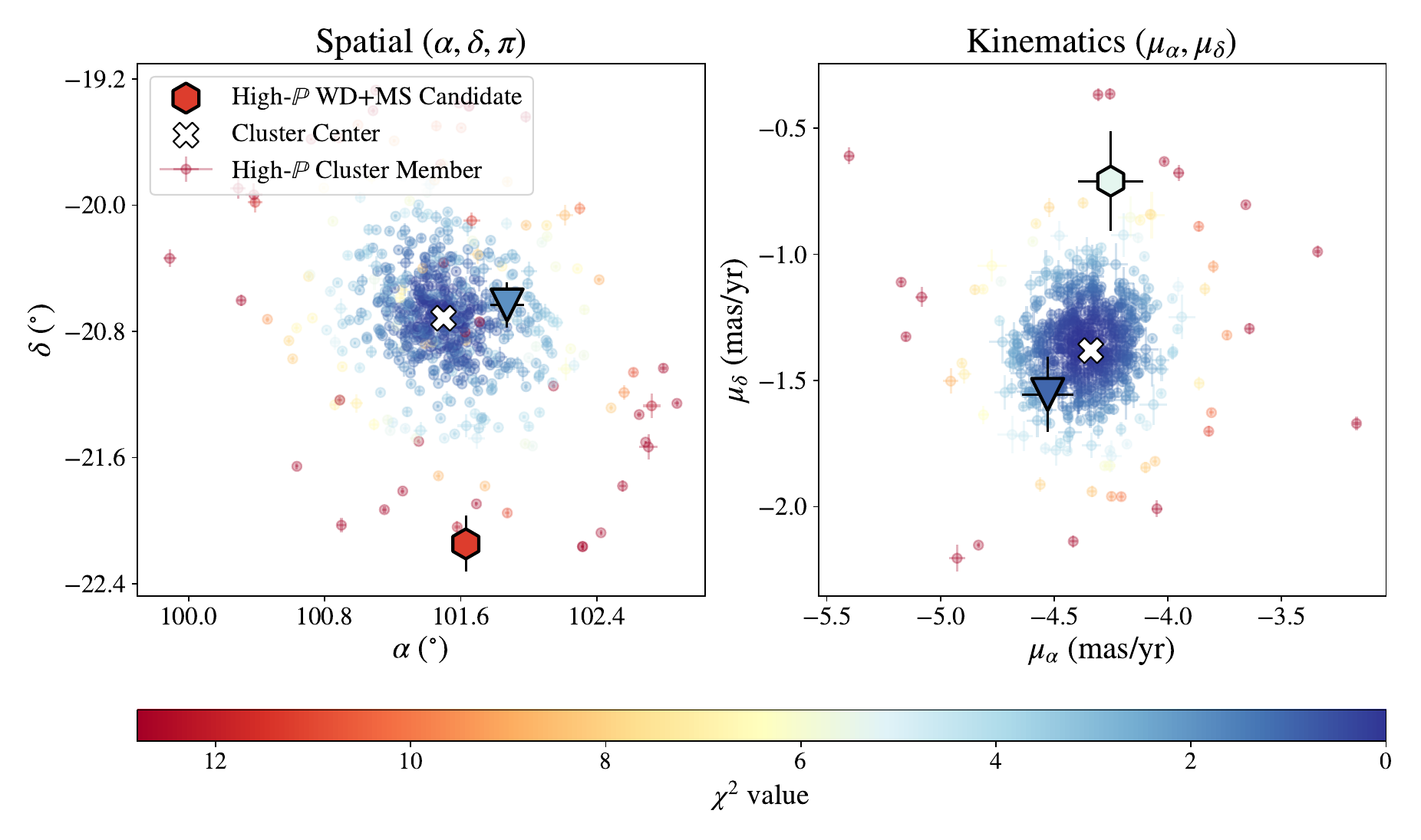}
    \caption{A 3-D spatial and 2-D kinematic $\chi^{2}$ association for two candidate WD+MS binaries in NGC 2287. NGC2287-c1 (triangle) and NGC2287-c2 (hexagon) are shown relative to high-probability cluster members in \cite{2020A&A...640A...1C}. Each source is colored by its $\chi^{2}$ value, where the spatial $\chi^{2}$ is computed using right ascension ($\alpha$), declination ($\delta$) and parallax ($\pi$), whereas the kinematic $\chi^{2}$ is computed from proper motions ($\mu_{\alpha}$, $\mu_{\delta}$). Both of these candidates fall into the `Tier 1' category for cluster membership, since they are both spatially and kinematically consistent with their host cluster.}
    \label{fig:kinematics}
\end{figure*}

In Sections \ref{sec:searchradius} and \ref{sec:kinematicclean}, we describe the initial spatial and kinematic constraints ensuring each candidate is at least broadly consistent with the properties of its host cluster. We performed those cuts in order to minimize computational expense, but chose liberal restrictions in order to allow for slight variations in both spatial location and proper motion compared to the high-probability cluster members identified by \cite{2020A&A...640A...1C}. Here, we perform a $\chi^{2}$ analysis to quantitatively determine the degree of spatial and kinematic cluster association for each candidate. A similar technique has previously been used  to identify possible foreground dwarfs in the direction of the Magellanic Clouds based on their \textit{Gaia} astrometric parameters \citep[e.g][]{Gaia2018,Ogrady2020,Neugent2020}. 

To begin, we must define a set of distributions that describe the spatial and kinematic properties for each of the OCs in our sample. We do not perform our own cluster membership analysis, but instead select the stars with cluster membership probabilities above $P>0.5$ from \cite{2020A&A...640A...1C}. We then use the \textit{Gaia} DR3 astrometric measurements for these stars to construct two separate covariance matrices (C) for each cluster: one that describes the 3-D spatial distribution  ($\overrightarrow{\mu}$ = [$\alpha$, $\delta$, $\pi$]) and one that describes the 2-D kinematic distribution ($\overrightarrow{\mu}$ = [$\mu_{\alpha}$, $\mu_{\delta}$]). Following \cite{Ogrady2023}, we take measurement uncertainties into account by minimizing the total negative log-likelihood for a set of matrices, each weighted by the measurement uncertainties of a single object. This yields an ``optimal'' covariance matrix, C$_{*}$, which is used in the rest of our analysis.  

For each of our candidate WD+MS systems, we then calculate a $\chi^{2}$ statistic between both its spatial and kinematic properties and those of the high-probability cluster members as: $\chi^{2} =$ $( \overrightarrow{\mu}-\overrightarrow{X} )^{T}\mathrm{C}_{*}^{-1}(\overrightarrow{\mu}-\overrightarrow{X})$, where $\overrightarrow{X}$ represents the median spatial or kinematic properties of the known OC members. We then categorize our candidates into three tiers based on their 3-D spatial ($\alpha, \delta, \pi$) and 2-D kinematic ($\mu_{\alpha}, \mu_{\delta}$) $\chi^{2}$ values. To consider a given candidate to be spatially or kinematically ``consistent'' with a given cluster we set thresholds of $\chi^{2} < 12.8$ and $\chi^{2} < 10.6$ for our 3-D spatial and 2-D kinematic analyses, respectively. Candidates with $\chi^{2}$ values that exceed these thresholds fall outside the region that contains 99.5\% of the high-probability cluster members in \cite{2020A&A...640A...1C}, making cluster association less certain. Our three tiers of candidates are summarized as follows:

\begin{enumerate}
\setlength\itemsep{-0.1em}
    \item \textit{Tier 1:} Candidates are both spatially and kinematically consistent with their host cluster. 22/52 high-probability WD+MS candidates ($\sim$ 42\%) fall in Tier 1. In our catalog, Tier 1 candidates are flagged with \texttt{Tier=`1'} and are the most secure cluster members in our sample (see Table \ref{candsnippet}).
    \item \textit{Tier 2:} Candidates are kinematically consistent with, but spatially offset from the high-probability cluster members in \citet{2020A&A...640A...1C}. 28/52 high-probability WD+MS candidates ($\sim$54\%) fall in Tier 2. Tier 2 candidates are flagged with \texttt{Tier=`2'} in our catalog. We explore the possibility that the inconsistent spatial locations for many of these systems could be due to natal kicks in Section \ref{sec:natal}. 
    \item \textit{Tier 3:} Candidates are both spatially and kinematically inconsistent with the host cluster. 2/52 high-probability WD+MS candidates ($\sim$4\%) fall in Tier 3. Tier 3 candidates are flagged with \texttt{Tier=`3'} in our catalog. While these candidates are located in the general vicinity of OCs--both spatially and kinematically---we caution that their cluster membership is less secure.
\end{enumerate} 

Figure \ref{fig:kinematics} shows an example of this $\chi^{2}$ analysis for the cluster NGC 2287, which contains two high-probability candidate WD+MS systems. Both NGC2422-c1 and NGC2422-c2 are classified as `Tier 1' WD+MS candidates, since their spatial locations and kinematics are consistent with a set of 625 high-probability ($P>0.5$) cluster members from \cite{2020A&A...640A...1C}.

Zero candidates in our sample have consistent spatial locations but inconsistent kinematics with respect to the host cluster. It is, however, worth noting that some OCs in our sample are diffuse. Specifically, three clusters (ASCC 97, Coin-Gaia 20 and UPK 418) have less than 40 high-probability cluster members in \cite{2020A&A...640A...1C}. This limited number of stars leads to less certainty in the overall cluster properties, making associations for these clusters more uncertain. Moreover, seven of our high-probability WD+MS candidates (Basel11a-c1, Coin-Gaia20-c1, LP1800-c1,  NGC2168-c1, NGC2301-c1, NGC2323-c1 and UPK418-c1) have \textit{Gaia} parallax values that are less than three sigma ($\sigma_{\pi}/ \pi > 0.33$). While errors are taken into account when calculating the 3-D spatial $\chi^{2}$ values, these candidates would also have higher uncertainties when spatially associating them with a host cluster.

\begin{table*}
  \centering
  \caption{Candidate WD+MS Systems in Milky Way Open Clusters}\label{candsnippet}
  \begin{tabular}{ccccccccccc}
  \hline
  \hline
    No. & Candidate & \textit{Gaia} DR3 ID & $\alpha \ (^{\circ})$ & $\delta \ (^{\circ})$ & $g_{mag}$& ... & $\mathbb{P}$(WDMS)  & $\chi^{2} (\alpha, \delta, \pi)$ & $\chi^{2} (\mu_{\alpha}, \mu_{\delta})$ & Tier \\
    \hline
    1 & Alessi31-c1 & 4149442705819050752 & 265.63887 & -13.98342 & 19.941 & ... & 0.978 & 43.68 & 3.55 & 2\\
    2 & Alessi44-c1 & 4291533250606570752 & 295.85724 & 5.73953 & 19.295 & ... & 0.977 & 16.13 & 1.00 & 2\\
    3 & Alessi44-c2 & 4241644456689125504 & 295.65824 & 2.46757 & 20.677 & ... & 0.937 & 0.98 & 0.36 & 1\\
    4 & Alessi44-c3 & 4291030803855467392 & 294.33365 & 4.80986  & 20.459 & ... & 0.989 & 17.35 & 3.25 & 2\\
    
    \vdots &  &  & &  &  &  & \\
    49 & UPK18-c1 & 4101661538264736512 & 282.60819 & -15.46526 & 19.300 & ... & 0.995 & 46.94 & 6.23 & 2\\
    50 & UPK305-c1 & 335055179059807104 & 39.15801 & 39.66823 & 18.664 & ... & 0.927 & 1.91 & 0.53 & 1\\
    51 & UPK418-c1 & 3138459356262913280 & 115.43704 & 5.04302 & 21.023 & ... & 0.956 & 30.94 & 0.29 & 2\\
    52 & UPK442-c1 & 2951386008373222144 & 96.40236 & -14.11618 & 20.353 & ... & 0.985 & 12.07 & 0.29 & 1\\
    \hline
    \hline
  \end{tabular}
  \tablenotetext{\dagger}{This table represents an example subset of our candidate WD+MS binary catalog.} 
  \vspace{0.1in}
\end{table*}

\subsection{Final Candidate catalog}\label{sec:finalcands}

As described previously, we photometrically identify 52 high-probability candidate WD+MS binaries in 38 OCs with our SVM, thus composing our catalog. Table \ref{candsnippet} highlights an example subset of systems and parameters included in this catalog. A full list of catalog parameters is described below.

\begin{itemize}
    \item \textit{Gaia DR3}: \textit{Gaia} identifiers, astrometry ($\alpha$, $\delta$, $\pi$, $\mu_{\alpha}$, $\mu_{\delta}$), photometric bands ($G$, $G_{BP}$, $G_{RP}$), quality indicators (\texttt{ruwe},  astrometric excess noise) and all associated uncertainties.

    \item \textit{Pan-STARRS1:} photometric bands ($g$, $r$, $i$, $z$, $y$) and associated uncertainties.

    \item \textit{2MASS}: 2MASS identifiers, photometric bands ($J$, $H$, $K_{s}$) and associated uncertainties.
    
    \item \textit{Model outputs:} SVM probabilities, spatial and kinematic $\chi^{2}$ values, and kinematic tiers.
\end{itemize}

\section{Candidate Follow-Up Observations} \label{sec:verification}

Our catalog contains 52 high-probability WD+MS binary candidates in 38 OCs, whose broadband photometric properties are consistent with those of SDSS WD+MS binaries identified by \cite{2010MNRAS.402..620R}. While spectroscopic follow-up of the full sample and detailed cases studies of individual objects will be presented in future works, here we present spectroscopy and high-cadence light curve analysis for a few example systems in order to provide context on their nature. Discussion of previous classifications for objects in our sample as well as other possible origins will be provided in Section \ref{sec:discussion} below. \newpage

\subsection{Spectroscopic Follow-Up}\label{sec:spectra}

\begin{figure*}
    \centering
    \includegraphics[width=1\textwidth]{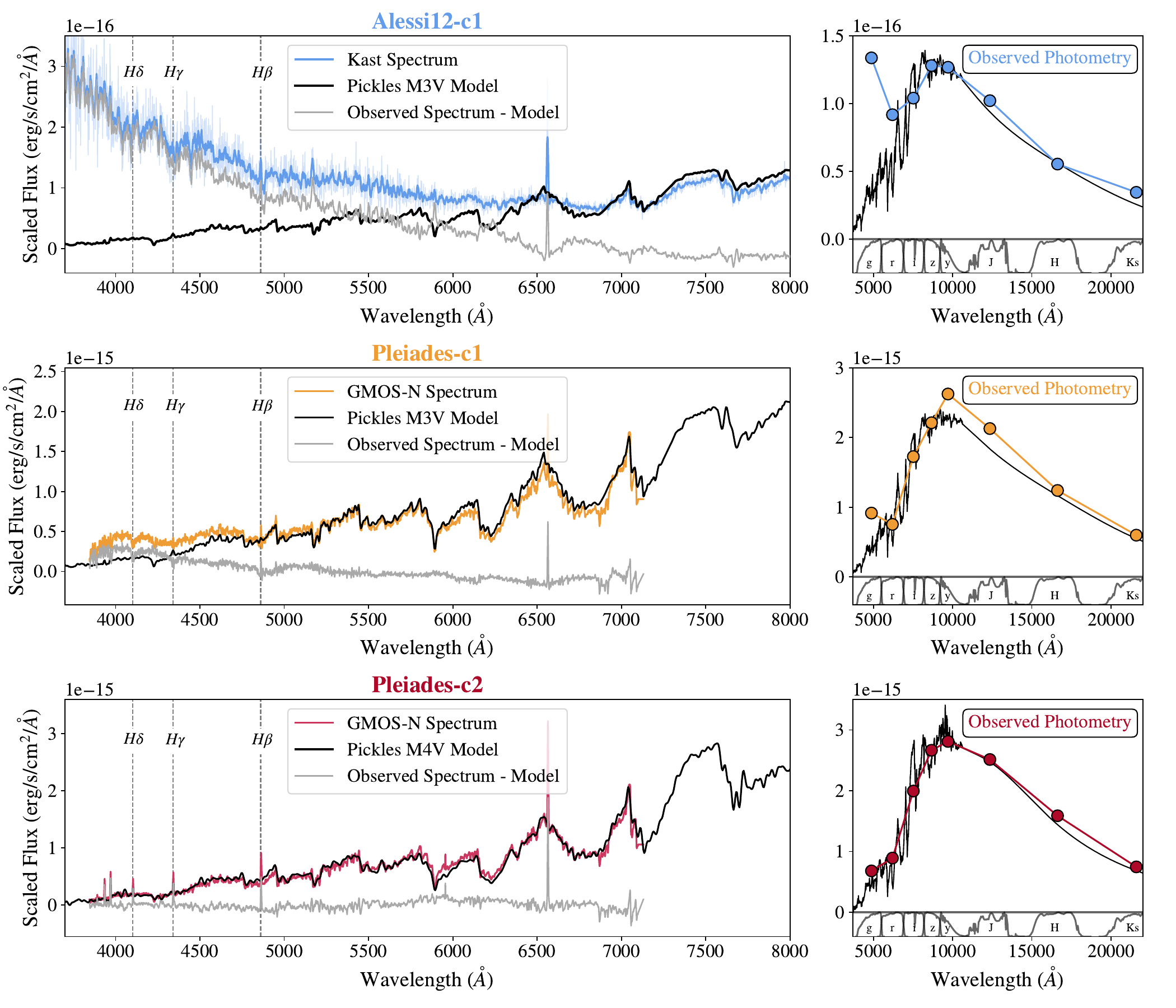}
    \caption{An example set of spectra (left panels) and observed photometry (right panels) for three high-probability candidate WD+MS binaries in our catalog. For each system, we show an observed spectrum/photometry (colored), a best-fit \cite{1998PASP..110..863P} M-dwarf model (black) and a residual spectrum (gray). Photometric bands for Pan-STARRS1 and 2MASS along with Balmer absorption features ($H\beta$, $H\gamma$, $H\delta$) are plotted for reference. \textit{Top:} Alessi12-c1 (the \textit{bluest} source in our catalog); this observed spectrum was obtained with the Kast spectrograph at the Lick Observatory. For reference, we plot both the raw and smoothed spectrum.  A large blue continuum flux excess and clear Balmer absorption features are present, indicating the presence of a WD in the system.  At redder wavelengths, Alessi12-c1 nicely fits the shape of a M3V star. Hence, Alessi12-c1 is a clear WD+MS binary. \textit{Middle:} Pleiades-c1 (the \textit{reddest} source in our catalog); this observed spectrum was obtained with the GMOS at the Gemini-North Telescope. While less dramatic than Alessi12-c1, Pleiades-c1 also exhibits an excess blue continuum flux and broad Balmer absorption features in its spectrum. While there is a slight infrared excess, Pleiades-c1 matches a M3V star at redder wavelengths. \textit{Bottom:} Pleiades-c2; like Pleiades-c1, this observed spectrum was obtained with the GMOS at the Gemini-North Telescope. Neither Balmer absorption lines nor a blue continuum flux excess are observed, however both the spectrum and photometry match that of an M4V Pickles model star.}
    \label{fig:spectra}
\end{figure*}

To investigate the types of systems in our catalog, we perform follow-up spectroscopy on a subset of our high-probability candidate WD+MS binaries. Based on their relative masses and temperatures, WD+MS binaries can exhibit a variety of spectral features, so we observe systems that span a range of colors ($1.0 < \text{BP-RP} < 2.6$) to probe different system classes. To yield higher signal-to-noise spectra, we also prioritize observing systems that are relatively bright ($m_{g}<19$ mag). In Figure \ref{fig:spectra}, we present spectra and SEDs for three high-probability candidates (Alessi12-c1, Pleiades-c1 and Pleaides-c2). These objects are labeled with numbers 1, 2 and 3, respectively in Figure~\ref{fig:highprobcandidates} and were chosen to highlight the types of systems identified in this study. Details of the observations and resulting spectra are described below. 

\subsubsection{Data Acquisition and Reduction}

We observed Alessi12-c1 using the Kast spectrograph mounted on the 3-meter Shane telescope at Lick Observatory \citep{millerstone93} over the course of several nights. Our instrumental setup used a 2$\arcsec$ slit, a D57 dichroic beamsplitter, the 600/4310 grism on the blue side of the spectrograph, and a 600/7500 grating on the red side. This provided a resolution of $\sim$5~\AA\ over the range of 3520--8750~\AA. Individual exposure times were 1830-seconds and 900-seconds on the blue and red sides, respectively. The data reduction process followed standard procedures described by \citet{silverman2012}.
In the top panel of Figure~\ref{fig:spectra}, we present the combined data taken over 2.5 hours in cloudy conditions on 2023 October 24 UT.

Observations of Pleiades-c1 and Pleiades-c2 were obtained using the Gemini Multi-Object Spectrograph (GMOS; \citealt{GMOS2004}) on Gemini-North as part of our program GN-2022A-Q-123 (Principle Investigator: Steffani M. Grondin). For both objects, we used a 1$\arcsec$ slit, the B600 grating and multiple subexposures were taken at two central wavelengths (540 and 550 nm) in order to cover the GMOS chip gaps. Overall, this provides a resolution of $\sim$5~\AA\ over the range of $\sim$3800-7100 \AA. All observations were obtained at the parallactic angle. Data was reduced using standard \texttt{gemini gmos} packages in \texttt{pyraf}. This includes bias subtraction, flat field correction, source extraction and wavelength calibration. Flux calibration was also performed using a combination of \texttt{pyraf} tasks and observations of spectrophotometric standards taken in the same set-up as our observations on a different night. Final spectra of Pleiades-c1 and Pleiades-c2 are shown in the lower two panels of Figure~\ref{fig:spectra}.

\subsubsection{Description of Observed Properties}

In Figure~\ref{fig:spectra} we present both the optical spectra (left panels) and optical-to-infrared SEDs (right panels) of Alessi12-c1, Pleiades-c1, and Pleiades-c2. Also shown are representative Pickles \citep{1998PASP..110..863P} models of M-type MS stars and the residuals after subtracting the fiducial Pickles model from the observed spectrum. As described above, these three objects were chosen to span the range of observed properties of our sample and include both the bluest (Alessi12-c1) and reddest (Pleiades-c1) objects in our final sample (in terms of their \textit{Gaia} BP-RP color; see Figure~\ref{fig:highprobcandidates}).

At a high level, there are several features that appear in the observed spectra of all three objects. First, they all show clear molecular absorption lines at redder wavelengths, indicative of an M-type star. Second, all three objects show narrow hydrogen Balmer lines in emission. Such features have previously been observed in close WD+MS binares (often attributed to the heated face of the secondary star e.g. \citealt[][]{Parsons2012}) and are also common in young/active M-dwarfs (caused by chromospheric activity e.g. \citealt[][]{2017ApJ...834...85N}).

However, clear differences are also evident, particularly at bluer wavelengths.
Alessi12-c1 is an example of a system that is clearly a WD+MS binary; a large blue continuum flux excess and clear Balmer absorption features are observed in the blue, while the spectrum matches a Pickles M3V stellar model at redder wavelengths. This spectrum is in agreement with the photometry of Alessi12-c1, which nicely fits the shape of a M3V star in the $i$ to $K$ bands, but has a significant flux uptick in the Pan-STARRS1 $g$ and $r$ bands.

While significantly less dramatic than Alessi12-c1, Pleiades-c1 (which was the \emph{reddest} object in our sample) also exhibits an excess blue continuum flux and broad Balmer absorption features. The spectrum of Pleiades-c1 also matches that of an M-dwarf at redder wavelengths, however Pleiades-c1 has a slight infrared excess compared to a Pickles M3V model star (see Section \ref{sec:yso}). Similar to Alessi12-c1, the photometry of Pleaides-c1 reveals a clear flux uptick in the $g$-band compared to the M3V star. We note that this object was also previously identified as a spectroscopic binary (see Section~\ref{sec:previous-class}).

Finally, we discuss Pleiades-c2. This object has a similar, but slightly bluer, \textit{Gaia} $BP-RP$ color to Pleiades-c1 and is located similarly close to the ZAMS in Figure~\ref{fig:highprobcandidates}. However, as seen in the lower right panel of Figure~\ref{fig:spectra}, while a small $g$-band deviation from the M4V dwarf model is observed (likely leading to its selection by our SVM model) this is much more subtle than observed even in Pleiades-c1. In addition, when we compare our observed spectrum to the M4V model, the only significant deviation observed is the narrow Balmer emission lines described above. In particular, the spectrum does not show clear evidence for \emph{either} broad Balmer absorption lines or a continuum flux excess down to wavelengths of $\sim$4000 \AA. While this object may represent a case with very low flux contribution from the WD in the optical (which would then be better served by ultraviolet follow-up) it is also possible that it is a (single) active M-dwarf whose Pan-STARRS1 g-band photometry was impacted by flares. We will further address whether this is expected to be a significant contaminant in our sample in Section~\ref{sec:Mdward-contamination}, below.

Detailed modelling and RV follow-up for Alessi12-c1, Pleaides-c1, Pleiades-c2, and other objects in our final sample are ongoing and will be presented in future work. We emphasize again, that these three objects were chosen as representative examples. SEDs for all 52 candidates are provided in Appendix \ref{sec:seds}, where it is clear that other sources in our catalog have similar photometric properties to each of those discussed in detail here.

\subsection{Light Curve Data}

\begin{figure*}
    \centering
    \includegraphics[width=\textwidth]{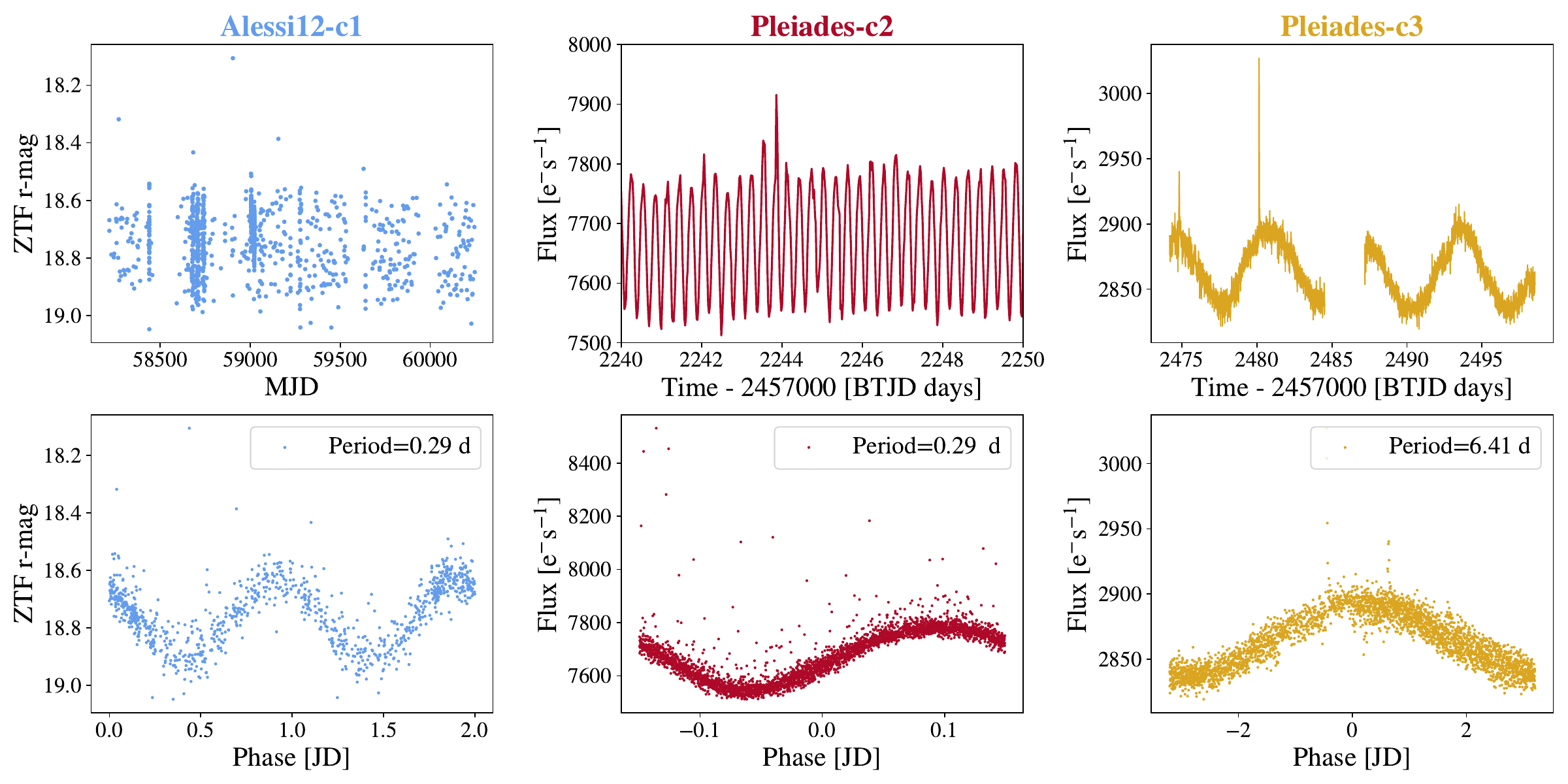}
    \caption{Raw and phase-folded light curves for 3 high-probability WD+MS binary candidates. \textit{Left panels:} Alessi12-c1; r-band light curve obtained from the Zwicky Transient Facility. The period of 0.29 days is computed using a a Lomb-Scargle periodogram where the phase folded light curve corresponds to the period of maximum power in the periodogram. \textit{Middle panels:} Pleiades-c2; light curve obtained from \textit{Kepler/K2} archival data. The period of 0.29 days is computed from a periodogram with default \texttt{LightKurve} settings. \textit{Right panels:} Pleiades-c3; light curve obtained from \textit{TESS} archival data. The period of 6.41 days is also computed from a periodogram with default \texttt{LightKurve} settings.}
    \label{fig:lightcurves}
\end{figure*}

Although more ambiguous than spectroscopy, observed system variability can provide context for the nature of an astronomical system. While light curves containing eclipses are the strongest variability indicators of both the presence of a binary and its period, eclipsing binaries are unfortunately rare. However, light curves exhibiting clear, short-period variability could indicate the presence of ellipsoidal modulations, which occur due to tidal distortions of stars in tight binaries. In addition, sinusoidal variability can be observed at a variety of timescales due to the rotation of M-dwarfs. When located in a close binary, they may be expected to be tidally-locked \citep[e.g.][]{Muirhead2013}, leading to shorter periods (although rotation periods of $\lesssim$ 1 day are also observed in young, single M-dwarfs e.g. \citealt{2017ApJ...834...85N}).
In practice, it is highly difficult to disentangle variability from binary ellipsoidal modulations and rapidly rotating M-dwarfs without radial velocity monitoring. However, since light curves with short-timescale variability can identify targets which may be short-period binaries (as expected for post-CE systems), we search for variability of high-probability candidates in our catalog.

To identify possible variability, we perform an archival search for light curve data of each high-probability WD+MS candidate in the \textit{Kepler/K2} Mission \citep{2014PASP..126..398H}, Transiting Exoplanet Survey Satellite \citep[\textit{TESS},][]{2015JATIS...1a4003R} and Zwicky Transient Facility \citep[ZTF,][]{2019PASP..131a8003M} databases. We query \textit{Kepler/K2} and \textit{TESS} light curves using the \texttt{Python} package \texttt{Lightkurve}\footnote{ \url{https://github.com/lightkurve/}} \citep{2018ascl.soft12013L} and ZTF directly through the NASA/IPAC Infrared Science Archive API\footnote{\url{https://irsa.ipac.caltech.edu/docs/program_interface/ztf_lightcurve_api.html}}. In Figure \ref{fig:lightcurves}, we present both raw and phase folded light curves for three high-probability WD+MS binary candidates (Alessi12-c1, Pleiades-c2 and Pleiades-c3). Alessi12-c1 and Pleiades-c2 were also shown in the spectroscopy section above. Pleiades-c3 is another candidate with slightly bluer colors, which is labeled with the number ``4'' in Figure~\ref{fig:highprobcandidates}.

Other objects in our catalog also have light curve matches within the \textit{Kepler/K2}, \textit{TESS}, and ZTF---although we note that \textit{K2/TESS} matches are less common due to the combination of limited spatial coverage by \textit{Kepler/K2} and the relatively shallow depth of \textit{TESS} (over $90\%$ of our high-probability candidates have magnitudes fainter than 18th mag). Of the candidates with matches, the light curve properties are diverse, ranging from (i) clear short period ($\sim$ days) variability, (ii) clear longer period ($\sim$weeks to months) variability, (iii) irregular or no obvious variability, and (iv) limited coverage or poor quality photometry. Here, we present three representative examples of objects with high-quality photometry that show short-period variability, as this is expected for post-CE systems -- which are the ultimate goal of our study.

When data from multiple instruments or epochs are available, we analyze and present the light curve that visually appears to have the lowest systematic issues. Periods are estimated from a periodogram with default \texttt{LightKurve} settings (\textit{TESS/K2} data) or from the maximum power of a Lomb-Scargle periodogram calculated with \texttt{astropy}'s \citep{Astropy2022} \texttt{LombScargle} package (ZTF). Information for each of the three example light curves is presented below. 

\begin{enumerate}
\setlength\itemsep{-0.1em}
    \item \textit{Alessi12-c1}: This light curve was obtained from the ZTF catalog. When queried, Alessi12-c1 had 919 high-quality (catflags $< 32768$) $r$-band observations spanning $\sim$4.7 years. The median time between sequential observations is $\sim$30 minutes, with $\sim$64\% of observations taken within $<$1 day of another. This is driven primarily by a few high cadence observing campaigns between years 2 and 3 (see Figure~\ref{fig:lightcurves}; left panel). Outside these times, the median cadence between observations is two days. The estimated period is 0.29 days.
    \item \textit{Pleiades-c2}: This light curve was obtained from \textit{Kepler/K2} Campaign 4 (2015). Observations are taken at a cadence of 30 minutes and the estimated period is 0.29 days.
    \item \textit{Pleiades-c3}: This light curve was obtained from \textit{TESS} Sector 43 (2021) and represents a \textit{TESS} light curve from a full-frame image (TESS-SPOC). The observations are taken at a cadence of 10 minutes and the estimated period is 6.41 days. 
\end{enumerate}

We emphasize that determining the actual origin of variability (e.g. ellipsoidal modulations versus rotation) requires additional follow-up observations. However, we note that all three of these sources have short variability periods which, if correlated with a binary orbit (either $P_{\rm{orb}}$ for rotational variability or 0.5$\times P_{\rm{orb}}$ for ellipsoidal modulations), would be consistent with known eclipsing post-CE systems \citep{2010MNRAS.407.2362P}. For Pleiades-c2, which was the system that showed no clear signatures of a WD in its optical spectrum in Section~\ref{sec:spectra}, we highlight that a rotational period of $\lesssim$ 0.3 days, if not due to binarity, would indicate a very young, and thus likely active M-dwarf \citep[e.g.][]{2017ApJ...834...85N}.

\section{Discussion}\label{sec:discussion}

In the sections above, we developed a machine learning framework to identify candidate WD+MS binaries in Milky Way OCs based on their broadband photometry. In total, after vetting the quality of each candidate, we present a catalog of $\sim$50 high-probability candidates. Here, we discuss our catalog in the context of the three previously known WD+MS binaries in OCs (Section~\ref{sec:previous-OC}), previous classifications for a subset of our sample (Section~\ref{sec:previous-class}), other possible origins for objects in our catalog (Section~\ref{sec:contaminants}), and implications of the kinematics observed in our sample for natal kicks of WD+MS binaries (Section~\ref{sec:natal}).

\subsection{Other Known WD+MS Systems in OCs}\label{sec:previous-OC}

As described in Section~\ref{sec:intro}, there are currently only three other WD+MS binaries known in OCs. V471 Tau \citep{1970PASP...82..699N, 1971ApJ...166L..81Y, 2022AJ....163...34M} and HZ9 \citep{1987AJ.....94..996S} are the only two post-CE WD+MS binaries that have been associated with an OC. Additionally,  \cite{2019ApJ...880...75R} present a magnetic WD+MS binary in NGC\,2422. None of these three objects are contained within our final catalog. Here, we discuss why they were not identified.  

Both V471 Tau and HZ9 are located in the Hyades cluster, which we do not search in this study. The Hyades is the closest OC to Earth located at a distance of $\sim 50$ pc \citep{2020A&A...640A...1C}. Consequently, the Hyades spans a large area on the sky, where the radius containing half its cluster members is $\sim5.5^{\circ}$ \citep{2020A&A...640A...1C}. Hence, using our method to search for WD+MS binaries in the Hyades would require a very large search radius. Moreover, the Hyades has a proper motion of $\mu_{\alpha}, \mu_{\delta} = (103.018, -27.228)$ mas/yr \citep{2020A&A...640A...1C}, which could potentially lead to cross-matching issues between newer and older surveys. 

The magnetic WD+MS cluster binary identified in NGC 2422 by \cite{2019ApJ...880...75R} is an excellent candidate to probe CE evolution. While an orbital period is not yet measured for this system, spectroscopy indicates an obvious WD and its kinematics show that it is conclusively associated with an OC. NGC 2422 is examined in this study, however the apparent \textit{Gaia} magnitude of this binary $m_{G} = 19.8$ is below our magnitude threshold of $m_{G} < 19.5$. Hence, this source is not detected in this catalog, but represents a strong candidate for radial velocity follow-up observations. 

\subsection{Previous Classifications of Candidates}\label{sec:previous-class}

To further understand the types of systems identified in this catalog, we query all of our high-probability candidates in the SIMBAD astronomical database \citep{2000A&AS..143....9W}. Of our 52 high-probability candidates, only six sources have SIMBAD classifications, where the object types are described below. We note that none of our candidates have ever been classified as WD+MS binaries. We also examine how many of the candidates selected by our method were identified as likely cluster members in the catalog of \citet{2020A&A...640A...1C} (which we use as the kinematic baseline for our target selection in Section~\ref{sec:starsample}). \newpage

\subsubsection{Binary Systems}\label{sec:binarystudy}

With the goal of studying binary systems and multiples in young stellar clusters or associations, \cite{2019AJ....157..196K} used the Apache Point Observatory Galactic Evolution Experiment (APOGEE) spectrograph to target sources in the Orion Complex with infrared excess, optical variability and previous young stellar object (YSO) literature designations. This study also used cross-correlation functions to autonomously classify double-lined spectroscopic binaries (SB2s). In this study, Pleiades-c1 was found to be a SB2, indicating it is likely a binary system.

\subsubsection{Eruptive Variable Stars}\label{sec:eruptive}

Pleiades-c2 and Pleiades-c3 (designated `V754 Tau' and `V880 Tau'', respectively) are classified as eruptive variable stars in the General catalog of Variable Stars \citep[GCVS;][]{2017ARep...61...80S}. Originally identified in \cite{1982BITon...3....3H}, the GCVS abbreviation for these stars is `\texttt{UV}', which represent the class of `UV Ceti' type variable stars. UV Ceti stars are highly magnetic K or M-type dwarfs who undergo flaring on the order of seconds to minutes \citep[e.g.][]{1976ApJS...30...85L}.

\subsubsection{Young Stellar Objects}\label{sec:yso}

YSOs represent a broad class of objects, spanning protostars to pre-MS stars which are generally $\lessapprox 100$ Myr in age \citep[e.g.][]{2012RAA....12....1P}. One of our targets, Pleiades-c1, is listed as a YSO in SIMBAD and originally classified in \cite{2019AJ....157..196K}. As mentioned in Section~\ref{sec:binarystudy}, this paper obtained APOGEE spectra of a sample of over 5000 objects in the direction of nearby star-forming regions. Pleiades-c1 was selected for targeting in this paper based on the presence of an infrared excess in its SED.

Pleiades-c1 was specifically listed as a Class III YSO, which are systems that primarily show reddened blackbody SEDs in the NIR and weak excess emission at longer wavelengths \citep{Lada1987}. They are thought to be more evolved, diskless, YSOs during the later stages of their pre-MS evolution \citep[e.g.][]{2002EAS.....3.....B}. As mentioned in Section~\ref{sec:complete}, for clusters with ages $\lesssim$ 100 Myr (as is the case for the Pleiades) it is possible that some low mass stellar systems have not yet reached the MS. Pleiades-c1 may therefore be an example containing such an object. As described in Section~\ref{sec:spectra} and shown in Figure~\ref{fig:spectra}, its optical spectrum is well matched to an M3V-M4V model in the red with evidence for Balmer absorption lines from a binary companion in the blue and potential excess emission in the infrared.

\newpage
\subsubsection{Stars}

Pleiades-c4, NGC6568-c1 and NGC752-c1 are all assigned the general classifier `Star' in SIMBAD.

\subsubsection{Open Cluster Members}

Three high-probability candidates in the Pleiades (Pleiades-c1, Pleaides-c2 and Pleiades-c3) are listed as cluster members in the \cite{2020A&A...640A...1C} catalog. Each source is indicated to be a cluster member with a probability of $P=1.0$, which is consistent with the low spatial and kinematic $\chi^{2}$ values for these candidates measured in Section \ref{sec:kinematics-candidates}. NGC6568-c1 is also present in the \cite{2020A&A...640A...1C} catalog, with a cluster probability of $P=0.2$ (likely because NGC6568-c1 exhibits kinematics that are off cluster centre). These are the only high-probability WD+MS candidates in our catalog that are identified as cluster members in \cite{2020A&A...640A...1C}, however this is likely due to the fact that (i) all cluster members used in the \cite{2020A&A...640A...1C} catalog have an apparent \textit{Gaia} DR2 magnitude of $m_{G} < 18$ (whereas 48 of 52 high-probability candidates are fainter than this) and (ii) the large spatial cluster offsets for many of our systems could reduce membership probabilities.

\subsection{Other Possible Candidate Origins}\label{sec:contaminants}

\begin{figure*}
    \centering
    \includegraphics[width=\textwidth]{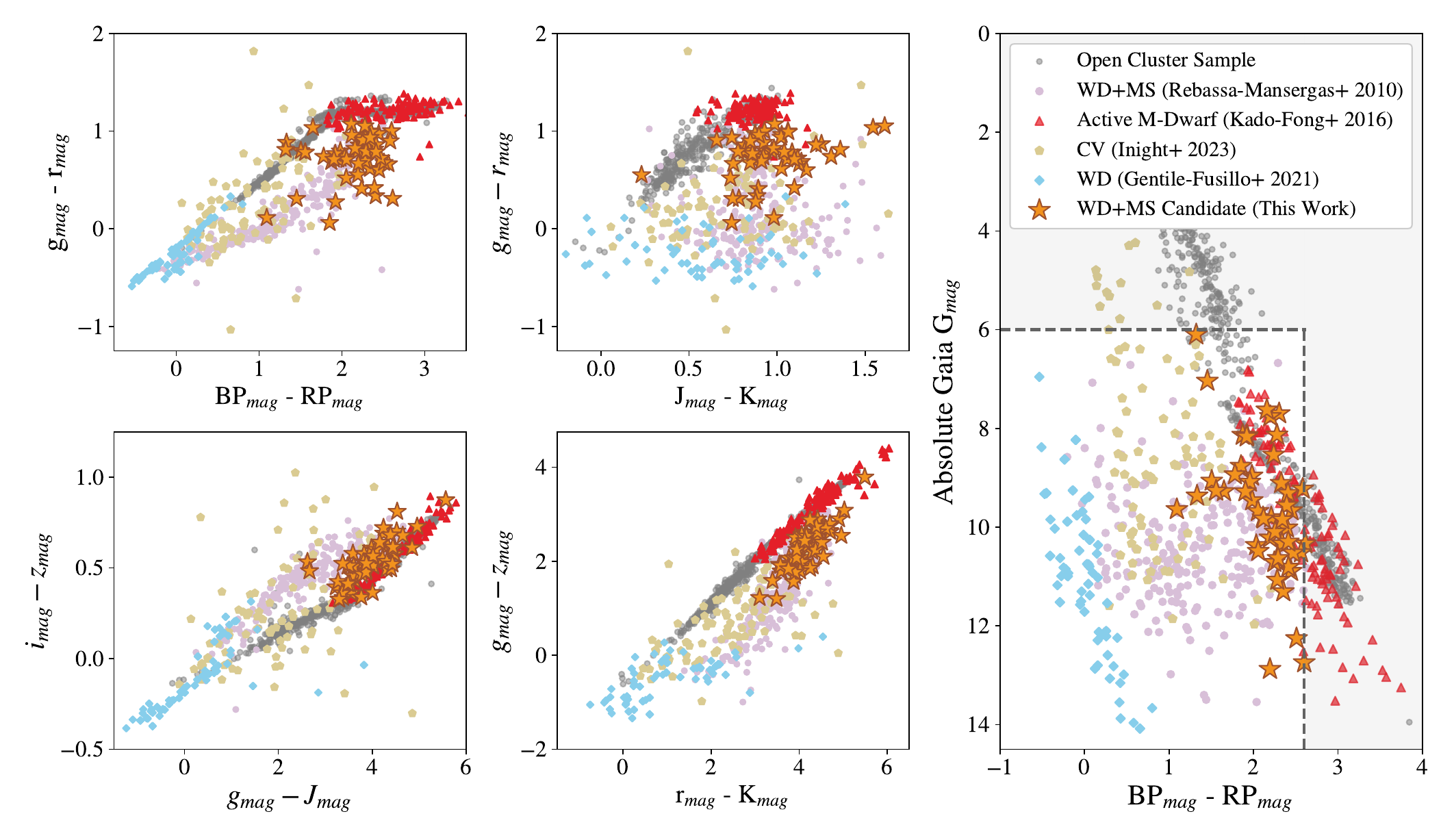}
    \caption{color-color and color-magnitude spaces of a variety of astrophysical sources with similar photometry to SDSS WD+MS binaries in \citet{2010MNRAS.402..620R}. Here, we show rapidly rotating M-dwarfs (red triangles) from \citet{2016ApJ...833..281K}, cataclysmic variables (gold hexagons) from \citet{2023MNRAS.524.4867I} and single WDs (blue diamonds) from \citet{2021MNRAS.508.3877G} relative to our high-probability candidate WD+MS binaries (orange stars) and our training samples (SDSS WD+MS binaries from \cite{2010MNRAS.402..620R} and a set of MS cluster members from \cite{2020A&A...640A...1C} as purple and gray circles, respectively). We conclude that our catalog is unlikely to have any significant contamination from single WDs, but our sample may contain CVs (in addition to detached WD+MS binaries). Some contamination from young, active M-dwarfs is possible for systems in our catalog that lie close the the ZAMS.}
    \label{fig:contaminants}
\end{figure*}

From our spectra in Figure \ref{fig:spectra}, it is clear that WD+MS binaries are present in our catalog. Since spectroscopy would be required to determine the nature of all systems, we investigate a variety of other possible astronomical sources that could be present in our sample. Each possible object is summarized below, where the respective photometric properties of each population are presented in a \textit{Gaia} CMD and four color-color spaces in Figure \ref{fig:contaminants}.

\subsubsection{Cataclysmic Variables}

Cataclysmic variables (CVs) are interacting binary star systems where a WD actively accretes matter from a low-mass MS companion star \citep{1995cvs..book.....W}. CVs represent a broad class of binaries (e.g. novae, polars, AM Canum Venaticorum stars, etc.) where the sub-classes predominantly arise from differences in the WD accretion disk and magnetic fields \citep[e.g.][]{2001PASP..113..764D}. While CVs are indeed WD+MS binaries (and have even likely undergone CE events to achieve their small orbital separations), active accretion makes them slightly different objects than the main targets of our overarching program. Nonetheless, CV photometry is likely similar to that of a detached WD+MS binary system. \newpage

To investigate how CVs are classified with our SVM, we utilize the \cite{2023MNRAS.524.4867I} sample of CVs observed in SDSS I to IV. \cite{2023MNRAS.524.4867I} identify 507 CVs, representing the largest sample of spectroscopically confirmed CVs in the optical regime. After cross-matching this catalog with 2MASS, \textit{Gaia} DR3 and Pan-STARRS1 and ensuring robust photometry (Section \ref{sec:kinematicclean} and \ref{sec:photclean}), 92 CVs remain in our final sample. 

From Figure \ref{fig:contaminants}, we see that despite being generally bluer and brighter, there is significant overlap with the \cite{2010MNRAS.402..620R} WD+MS binaries in a \textit{Gaia} CMD. As expected, the colors of CVs and WD+MS binaries are highly similar, where the CVs are mostly indistinguishable from the WD+MS binaries. These photometric similarities lead to $56\%$ of the CV sample being classified as high-probability WD+MS candidates. While the majority of our WD+MS candidates are redder than the CVs in a \textit{Gaia} CMD, follow-up spectroscopy would be beneficial to distinguish between detached and accreting WD+MS binaries.

\subsubsection{Main-Sequence+Main-Sequence Binaries}

In Figure \ref{fig:extinction}, we see that four binaries lie to the right of the ZAMS, which is a parameter space that could include binary systems composed of two MS stars. To investigate how MS+MS binaries are classified in our model, we compose a set of model binaries using  the \citet{1998PASP..110..863P} stellar atmosphere models described previously. From Figure \ref{fig:extinction}, we see that because we impose a brightness threshold of M$_{\rm{G}}>$ 6 mag, our search is only sensitive to single stars from late F to M type depending on the Galactic extinction. For both zero extinction ($A_{V}=0.0$) and high extinction ($A_{V}=3.0$), we randomly sample 1000 MS+MS binary configurations composed of F5V to M6V model stars.

We find that the MS+MS binaries almost exactly match the Pickles curves shown in Figure \ref{fig:extinction} for both extinction values. For binaries with hotter MS stars (e.g. F-type) and cooler companions (e.g. M-type), the relative flux ratios are such that the hotter star dominates the photometry, appearing as a single MS star. On the other hand, while similar mass combinations appear brighter in the \textit{Gaia} CMD, their colors are only minimally impacted. When run through our SVM model, zero MS+MS model binaries are classified as high-probability WD+MS systems for both the $A_{V}=0.0$ and $A_{V}=3.0$ cases, with the highest probability system being classified at $P=0.02$ (since these Pickles models were included in our training sample, it is unsurprising that they almost perfectly align with the MS photometry). Hence, we expect very little to no contamination from MS+MS binaries in our catalog.

\subsubsection{Rapidly Rotating M-dwarfs}\label{sec:Mdward-contamination}

A significant fraction of low-mass stars are highly (or even fully) convective. Strong convection can generate large magnetic fields in stellar chromospheres, causing immense stellar activity in the form of eruptions or flares \citep{2003NewAR..47...85V}. The General catalog of Variable Stars \citep{2017ARep...61...80S} lists more than twenty sub-classes of eruptive variables, broadly categorized by their spectral type, flare frequency/regularity and astrophysical origin. Of particular interest are flaring (active) M-dwarf stars, which predominantly erupt in the near-ultraviolet portion of the electromagnetic spectrum and can thus manifest as blue color excesses in M-dwarf photometry if a flare was active when observations were obtained \citep{2007ApJS..173..673W, 2024arXiv240217384L}.  While the exact mechanisms that drive stellar flaring are still an open question, observational studies of chromospheric activity in M-dwarfs are extensive \citep[e.g.][]{2016IAUS..320..128D,2016ApJ...833..281K,2017ApJ...834...85N, 2017ApJ...849...36Y, 2020ApJ...905..107M, 2020AJ....160..219F,2020ApJ...892..144R, 2023ApJ...955...24R}. 

Since flares in M-dwarfs can contribute excess blue flux to otherwise red optical photometry, we consider how active M-dwarfs are classified with our SVM. \cite{2017ApJ...834...85N} provide a catalog of M-dwarfs with detailed activity indicators, however these stars are all located within the Solar Neighbourhood and are thus too bright to have reliable (unsaturated) photometric measurements in Pan-STARRS1. Thus, we consider the \cite{2016ApJ...833..281K} catalog of M-dwarfs observed in the Pan-STARRS1 Medium-Deep Survey, since these M-dwarfs are located at further distances. \cite{2016ApJ...833..281K} identify 270 rotating M-dwarfs, measuring periods of 0.7 days $< P <$ 130 days. While no activity indicators are explicitly included in this catalog, \cite{2017ApJ...834...85N} observe that almost all M-dwarfs with rotational periods of $P < 10$ days are active. After cross-matching this catalog with 2MASS, \textit{Gaia} DR3 and Pan-STARRS1, we find 126 M-dwarfs in the \cite{2016ApJ...833..281K} sample with $P < 10$ days. 110 of these stars adhere to the same photometric and kinematic constraints described in Section \ref{sec:kinematicclean} and \ref{sec:photclean}. We adopt this sample of rapidly rotating M-dwarfs as a proxy for the expected properties of active M-dwarfs. 

From Figure \ref{fig:contaminants}, we see that active M-dwarfs (red triangles) lie directly along the MS of our training data in both a CMD and all color-color spaces. Thus, we find that only three systems ($3\%$) are classified as high-probabilty WD+MS candidates with our SVM. However, despite these low numbers, the WD+MS candidates that lie close the ZAMS (approximately 25-30\% of our final sample) do share some photometric properties with active M-dwarf stars. For these systems, we conclude that some contamination is possible (especially since single M-dwarfs are expected to be more prevalent than WD+MS binaries) and spectroscopy is required to distinguish binarity from potential flaring. In particular, we note (i) Pleiades-c1 is located in this region of parameter space, but is a spectroscopically confirmed binary system and (ii) both Pleiades-c2 and Pleiades-c3 were previously classified as flaring M-dwarfs (Section~\ref{sec:eruptive}). However, also note that the presence of M-dwarf flares is not mutually exclusive with a potential WD+MS binary, as post-CE systems with active M-dwarf secondary stars have been identified \citep[e.g.][]{Muirhead2013}.

\subsubsection{Single WDs}

Depending on system age, single WDs can have effective temperatures between $10,000K < T_{eff} < 40,000K$ \citep{2006ApJS..167...40E, 2021MNRAS.508.3877G}. Hence, it is possible that older, cooler and redder WDs could share similar photometry to a WD+MS binary. To investigate this potential overlap, we consider the \cite{2021MNRAS.508.3877G} catalog of WDs identified in \textit{Gaia} Early Data Release 3. While the full sample contains  $\sim 359,000$ WDs, we find that upon cross-matching, only 88 of these systems have 2MASS data. 

In Figure \ref{fig:contaminants}, we observe that these 88 single WDs fall bluewards of the previously investigated populations in both a \textit{Gaia} CMD and all color-color spaces. While single WDs appear to occupy extensions of the WD+MS binary population in each color-color space, overlap with WD+MS binaries is present especially when 2MASS photometry is included. Due to this overlap, $38\%$ of the WDs in this catalog are classified by our SVM as probable WD+MS binaries. However, none of our identified WD+MS candidates visually overlap with the single WDs in any parameter space, making it unlikely that any of our candidates are actually single WDs.

\begin{figure*}
    \centering
    \includegraphics[width=1\textwidth]{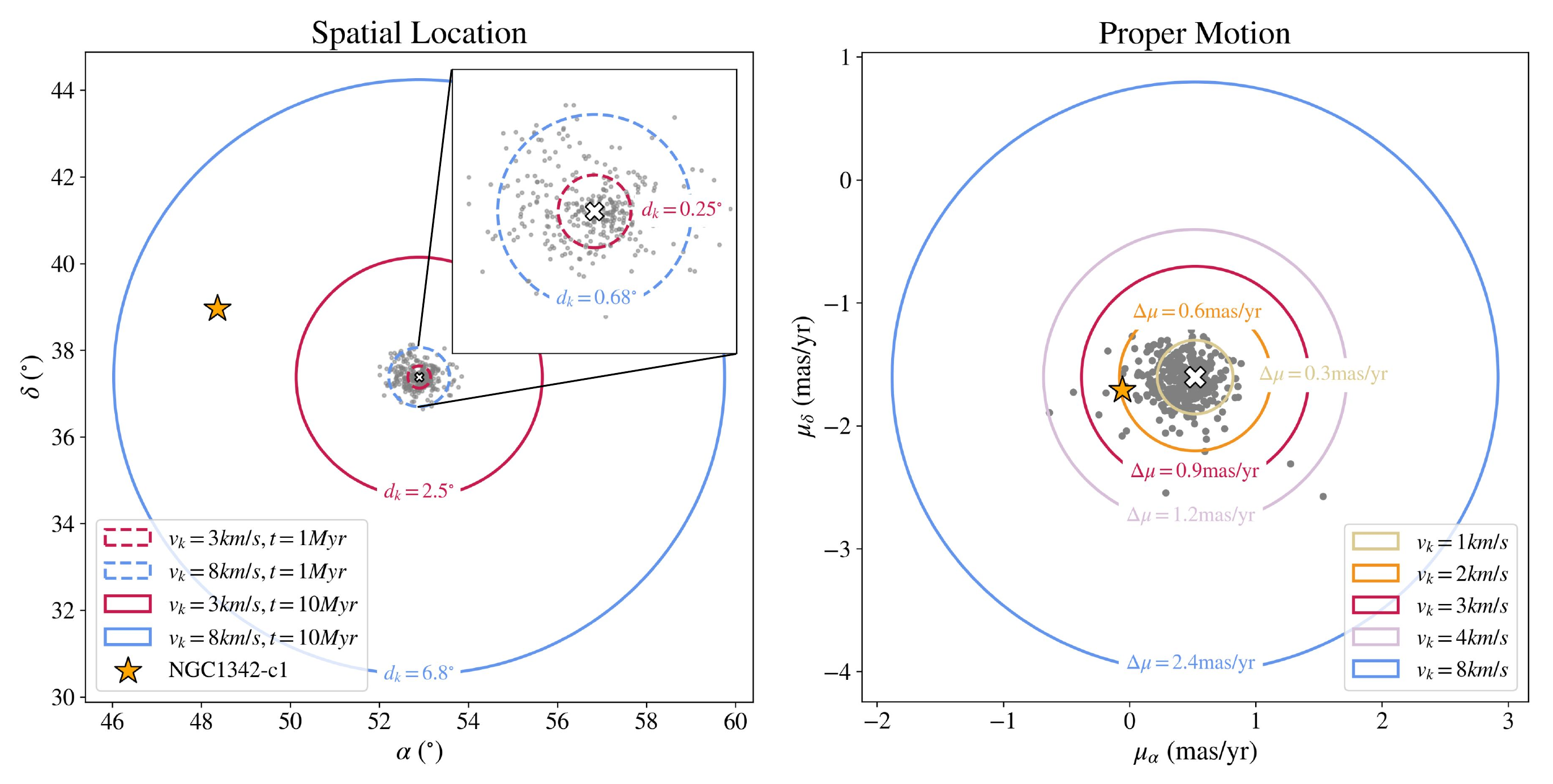}
    \caption{The proposed effect of a CE natal velocity kick on the spatial position and kinematics of candidate NGC1342-c1. Left: Spatial location of NGC1342-c1 (star) relative to high-probability cluster stars from \cite{2020A&A...640A...1C} (gray). Each circle represents the distance NGC1342-c1 would be located if it received a natal kick for various kick magnitudes and travel times. Right: Similar to the left panel, however with proper motion. Each circle here represents the change in NGC1342-c1's proper motion ($\Delta \mu$) upon receiving a velocity kick completely imparted into the tangential direction (for simplicity).}
    \label{fig:natalkicks2}
\end{figure*}

\newpage

\subsection{Implications for Natal Kicks}\label{sec:natal}

The observed number of WDs in OCs is far lower than expected \citep[e.g.][]{1977A&A....59..411W, 1992AJ....104.1876W, 2005ApJ...618L.129K, 2021ApJ...912..165R}. For decades, studies have postulated that cluster WD deficits could be caused by natal kicks ($v_{k}$) of a few km s$^{-1}$ received during the formation of the WD \citep{2003ApJ...595L..53F, 2007MNRAS.381L..70H, 2007MNRAS.382..915H, 2009ApJ...695L..20F}. Specifically, asymmetric winds of late-type asymptotic giant branch stars could result in asymmetric mass-loss, ultimately imparting a kick into a newly formed WD \citep{1993ApJ...413..641V}. Recent discoveries by \cite{2022ApJ...926..132H} and \cite{2022ApJ...926L..24M, 2023ApJ...956L..41M} have also identified WDs that likely escaped their host clusters. 

Similar to the case of single star evolution to a WD, 3-D hydrodynamical simulations of CE ejection \citep{1998ApJ...500..909S} have also found that mass can be asymmetrically ejected from the envelope, imparting natal kicks of $\sim3-8$ km s$^{-1}$ into post-CE binaries. These simulations are consistent with \cite{2010A&A...523A..10I} who find that natal kicks of $\sim$ a few km s$^{-1}$ allow for a better match between observed properties and models of barium stars (which could also be formed through CE evolution). Hence, it is possible that WDs in post-CE binary systems could experience multiple types of natal kicks during their evolution.

Upon examining the kinematics of the candidate WD+MS systems in our catalog, we find that while their overall proper motions are consistent with their host clusters, the spatial locations for more than \textit{half} of the candidates are largely offset (see Section~\ref{sec:kinematics-candidates}). Thus, we explore whether these spatial offsets could be explained by invoking a natal kick. As a fiducial example, we consider the WD+MS binary candidate NGC1342-c1. NGC1342-c1 has a proper motion consistent with being a member of NGC 1342 ($\chi^{2}=1.03$) but a spatial location inconsistent with the cluster ($\chi^{2}=112.41$) as computed in Section \ref{sec:kinematics-candidates}. 

To investigate its spatial offset, we estimate the distance NGC1342-c1 could travel if it was (i) indeed a WD+MS post-CE binary and (ii) originally located at the cluster centre. We assume two different kick velocities ($v_{k}=3$ km s$^{-1}$ and $v_{k}=8$ km s$^{-1}$, using the previously discussed post-CE natal kick range) and two travel times (1 Myr and 10 Myr). Coupling these kicks and travel times with the \cite{2020A&A...640A...1C} distance to NGC 1342 (686 pc), we find that NGC1342-c1 could have traveled between $0.25^{\circ} - 6.8^{\circ}$ from the cluster centre after ejection of its CE. The left panel of Figure \ref{fig:natalkicks2} shows that the current spatial position of NGC1342-c1 could be explained by these low velocity kicks and relatively short travel times.

We also consider how much the system's proper motion could change after experiencing a natal kick ($\Delta \mu$). In particular, we examine whether it would be reasonable for an object with a large spatial offset due to a natal kick to still have proper motions that overlap with the bulk of cluster members (as observed for most of the candidates in our sample). Again taking NGC1342-c1 as a fiducial case, we  determine $\Delta \mu$, by first using the distance and total proper motion of NGC1342 from \cite{2020A&A...640A...1C} to compute the system's pre-kick tangential velocity, $v_{t}$. Next, we compute the tangential velocity a system would have if it were kicked at $v_{k,t}=1$ km s$^{-1}$, 2 km s$^{-1}$, 3 km s$^{-1}$, 4 km s$^{-1}$ \ \text{and} \ 8 km s$^{-1}$. From here, we use $v_{t}$ and $v_{k,t}$ to compute $\Delta \mu$:

\vspace{-0.15in}
\begin{equation}
    \Delta \mu \text{[mas/yr]} = \frac{1000}{4.74\times d \text{[pc]}} \times (v_{k,t} - v_{t}) \text{[km s$^{-1}$]}
    \label{eq:2}
\end{equation}

The right panel of Figure \ref{fig:natalkicks2} highlights $\Delta \mu$ for a range of kick velocities. We see that the proper motion observed for NGC1342-c1 is consistent with expectations for a relatively low natal kick ($\sim2-3$ km s$^{-1}$). However, we note that---for simplicity---we assumed 100\% of the natal kick is received in the tangential direction of the velocity. Hence, these estimates of $\Delta{\mu}$ are likely \textit{overestimated}. This emphasizes that even larger kick velocities could yield a system with proper motions consistent with the host cluster. 

While specific cluster proper motions, distances and escape velocities determine the magnitude to which a natal kick might effect a post-CE system, this analysis shows that a post-CE system can be spatially offset from a cluster despite having a consistent proper motion. Hence, we conclude that it is possible for post-CE systems to migrate far from their birth cluster if a velocity kick is received. Importantly, we emphasize that $78\%$ of the OCs searched in this study revealed no WD+MS candidates. It is almost certain that there exist more WD+MS post-CE cluster binaries than presented in this catalog, however they are likely located outside the search radii chosen in this study. While extending search radii might help identify more WD+MS candidates, finding binaries that have travelled far beyond their host cluster is challenging. Although chemical tagging and dynamical simulations are good tools for associating escaped stars and binaries with individual star clusters \citep[e.g.][]{2023MNRAS.518.4249G, 2024MNRAS.528.5189G}, ample chemistry and radial velocities would be required for definitive associations of these escaped systems.

\section{Summary and Conclusions} \label{sec:summary}

In this paper, we search for WD+MS binaries in 299 OCs in the \cite{2020A&A...640A...1C} catalog. Using photometry from \textit{Gaia} DR3, 2MASS and Pan-STARRS1, we train a support vector machine model to identify systems that are photometrically similar to a sample of SDSS WD+MS binaries from \cite{2010MNRAS.402..620R}. Throughout, we design both our training sample and method to try to minimize contamination (and thus sacrifice some amount of completeness) in order to identify a set of high-probability candidates, which can then be followed-up for confirmation and characterization. While the focus of this study is identifying WD+MS binaries in clusters, we emphasize that this methodology could be utilized for photometric identification of a variety of other astrophysical sources. Figure \ref{fig:schematic} highlights the main steps in our identification method. Our main results are summarized below.

\begin{enumerate}
    \item This catalog contains 52 candidate WD+MS binaries in 38 OCs which were identified with high-probability ($P>0.9$) from our SVM classification model and subsequently vetted for data quality (Figure \ref{fig:highprobcandidates}). Generally, our catalog contains systems that are redder than the overall sample of WD+MS binaries presented in \cite{2010MNRAS.402..620R}. This is in line with other WD+MS systems identified in \textit{Gaia} \citep{2021MNRAS.504.2420I, 2021MNRAS.506.5201R}. 78\% of the 299 OCs searched had zero high-probability WD+MS binaries identified by our method.

    \item We present follow-up spectroscopy from the Lick and Gemini Observatories for three example candidates that span a range of \textit{Gaia} $BP-RP$ colors (Figure \ref{fig:spectra}). From this spectroscopy alone, it is clear that a variety of systems are present in this catalog. For instance, two objects (which represent both the bluest and reddest systems in our sample) are well-matched to M-type MS stellar models at red wavelengths, but show  evidence of broad Balmer lines and excess flux at blue wavelengths---as would be expected for a WD companion. In contrast, the third object shows no clear signatures of a WD in its optical spectrum. It may represent either a case with a limited flux contribution from the WD in the optical regime, or contamination in our sample from an active (flaring) M-dwarf. SEDs for all WD+MS candidate binaries are similar to the three examples shown. Hence, we expect that this catalog spans a wide range of WD and MS configurations.

    \item While our high-probability candidates exhibit a variety of light curve behavior, we show example \textit{TESS}, \textit{Kepler/K2} and ZTF light curves for three objects that show clear and regular variability with periods spanning $P=0.29 - 6.4$ days. This observed variability could either be caused by rapid rotation of an M-dwarf or ellipsoidal modulations from tidal distortion of a short-period binary.
    
    \item Since we solely use photometry to identify WD+MS candidates, we investigate the possibility that other astrophysical sources are present in our catalog. Specifically, we run populations of cataclysmic variables \citep{2023MNRAS.524.4867I}, MS+MS binaries \citep{1998PASP..110..863P}, rapidly rotating (active) M-dwarfs \citep{2016ApJ...833..281K} and single WDs \citep{2021MNRAS.508.3877G} through our SVM. Based on this and the locations of our candidate systems in the \textit{Gaia} CMD and various color-color spaces, we conclude that we are unlikely to have any significant contamination from single WDs or MS+MS binaries. In contrast, our sample may contain CVs in addition to detached WD+MS binaries, and we may have some contamination from young, active M-dwarfs in the $\sim25-30\%$ of our sample that lies close the the ZAMS. Ultimately, follow-up spectroscopy will be required to fully confirm the nature of each candidate.
    
    \item While the candidate proper motions are generally consistent with their host clusters, we find that over half of our candidates are spatially offset from their hosts. A possible explanation for this is a natal kick received during either the asymptotic giant branch phase or upon ejection of a CE.  We find that moderate velocity kicks between 3 km s$^{-1} < v_{k} < 8$ km s$^{-1} $ are sufficient to cause a large spatial offset, but retain consistent proper motions relative to the cluster (Figure \ref{fig:natalkicks2}). It is plausible that other WD+MS binaries exist for the clusters searched in this study, however they are likely located far beyond our chosen search radii.
\end{enumerate}

Ultimately, this catalog is a necessary first step in a larger effort to provide observational constraints on the CE phase and detailed characterization of a subset of candidates (e.g. Alessi12-c1, Pleiades-c1, etc.) identified in this study is currently underway. Once a larger sample of post-CE binaries in clusters is confirmed, pre-CE progenitor masses can be estimated, leading to a one-to-one mapping between the initial and final masses of systems that have undergone a CE event (similar to how the IFMR for isolated WDs is determined). With these observational benchmarks, this sample will aid in efforts to unlock important new insights into one of the most uncertain phases of binary evolution.

\section*{Acknowledgements} \label{sec:ack}

The authors thank Max Moe for providing insightful discussion about potential contaminant sources that aided in an improved understanding of the WD+MS candidates in our catalog. The authors also thank Alexander Laroche for useful input in validating our machine learning model and Fraser Evans and Phil Van-Lane for valuable edits in the manuscript. Helpful conversations with Samantha Berek, Victor Chan, Mairead Heiger, Marten van Kerkwijk, Gustavo Medina-Toledo and Jeremy Webb also greatly improved this research.

S.M.G. acknowledges the support of
the Natural Sciences and Engineering Research Council of Canada
(NSERC) and is partially funded through a NSERC Postgraduate
Scholarship – Doctoral. S.M.G. also recognizes funding from
a Walter C. Sumner Memorial Fellowship. M.R.D. acknowledges support from the NSERC through grant RGPIN-2019-06186, the Canada Research Chairs Program, and the Dunlap Institute at the University of Toronto. J.N. and P.M. acknowledge support from U.S. National Science Foundation award AST-2009713. J.N. also acknowledges support from U. S. National Science Foundation award AST-2319326.

This paper includes observations obtained at the international Gemini Observatory, a program of NSF NOIRLab, which is managed by the Association of Universities for Research in Astronomy (AURA) under a cooperative agreement with the U.S. National Science Foundation on behalf of the Gemini Observatory partnership: the U.S. National Science Foundation (United States), National Research Council (Canada), Agencia Nacional de Investigación y Desarrollo (Chile), Ministerio de Ciencia, Tecnología e Innovación (Argentina), Ministério da Ciência, Tecnologia, Inovações e Comunicações (Brazil), and Korea Astronomy and Space Science Institute (Republic of Korea). The Gemini data in this paper are from program GN-2022A-Q-123. This paper also includes data collected by the TESS mission. Funding for the TESS mission is provided by the NASA's Science Mission Directorate. This paper includes data collected by the Kepler mission and obtained from the MAST data archive at the Space Telescope Science Institute (STScI). Funding for the Kepler mission is provided by the NASA Science Mission Directorate. STScI is operated by the Association of Universities for Research in Astronomy, Inc., under NASA contract NAS 5–26555. ZTF is supported by the National Science Foundation under Grant
No. AST-2034437 and a collaboration including Caltech, IPAC, the Weizmann Institute for Science, the Oskar Klein Center at
Stockholm University, the University of Maryland, Deutsches Elektronen-Synchrotron and Humboldt University, the TANGO
Consortium of Taiwan, the University of Wisconsin at Milwaukee, Trinity College Dublin, Lawrence Livermore National
Laboratories, and IN2P3, France. Operations are conducted by COO, IPAC, and UW. This research has made use of the SIMBAD database, operated at CDS, Strasbourg, France. A major upgrade of the Kast spectrograph on the Shane 3~m telescope at Lick Observatory, led by Brad Holden, was made possible through generous gifts from the Heising-Simons Foundation, William and Marina Kast, and the University of California Observatories. Research at Lick Observatory is partially supported by a generous gift from Google.

\section{Data Availability}
Data from the Kepler/EPIC \citep{STScI2016-bw}, TESS Input Catalog and Candidate Target List \citep{https://doi.org/10.17909/fwdt-2x66}  and Pan-STARRS DR1 Catalog \citep{STScI2022-jj} were accessed from the Mikulski Archive for Space Telescopes (MAST) database.

\facilities{Gaia, Gemini (GMOS), Kepler/K2, Lick (Kast), Pan-STARRS, TESS, ZTF, 2MASS}

\software{\texttt{astropy} \citep{astropy:2013, astropy:2018, Astropy2022}, \texttt{lightkurve} \citep{2018ascl.soft12013L}, \texttt{matplotlib} \citep{Hunter:2007}, \texttt{numpy} \citep{harris2020array}, \texttt{pandas} \citep{mckinney-proc-scipy-2010}, \texttt{pyraf} \citep{2012ascl.soft07011S}, \texttt{scikit-learn} \citep{scikit-learn}, \texttt{TOPCAT} \citep{2005ASPC..347...29T}}

\bibliography{sample631}{}
\bibliographystyle{aasjournal}

\begin{appendix}\label{appendix:appendix}

\section{Photometric Search Examples}\label{ap:example-search}
\begin{figure*}[!ht]
    \centering
    \includegraphics[width=0.62\textwidth]{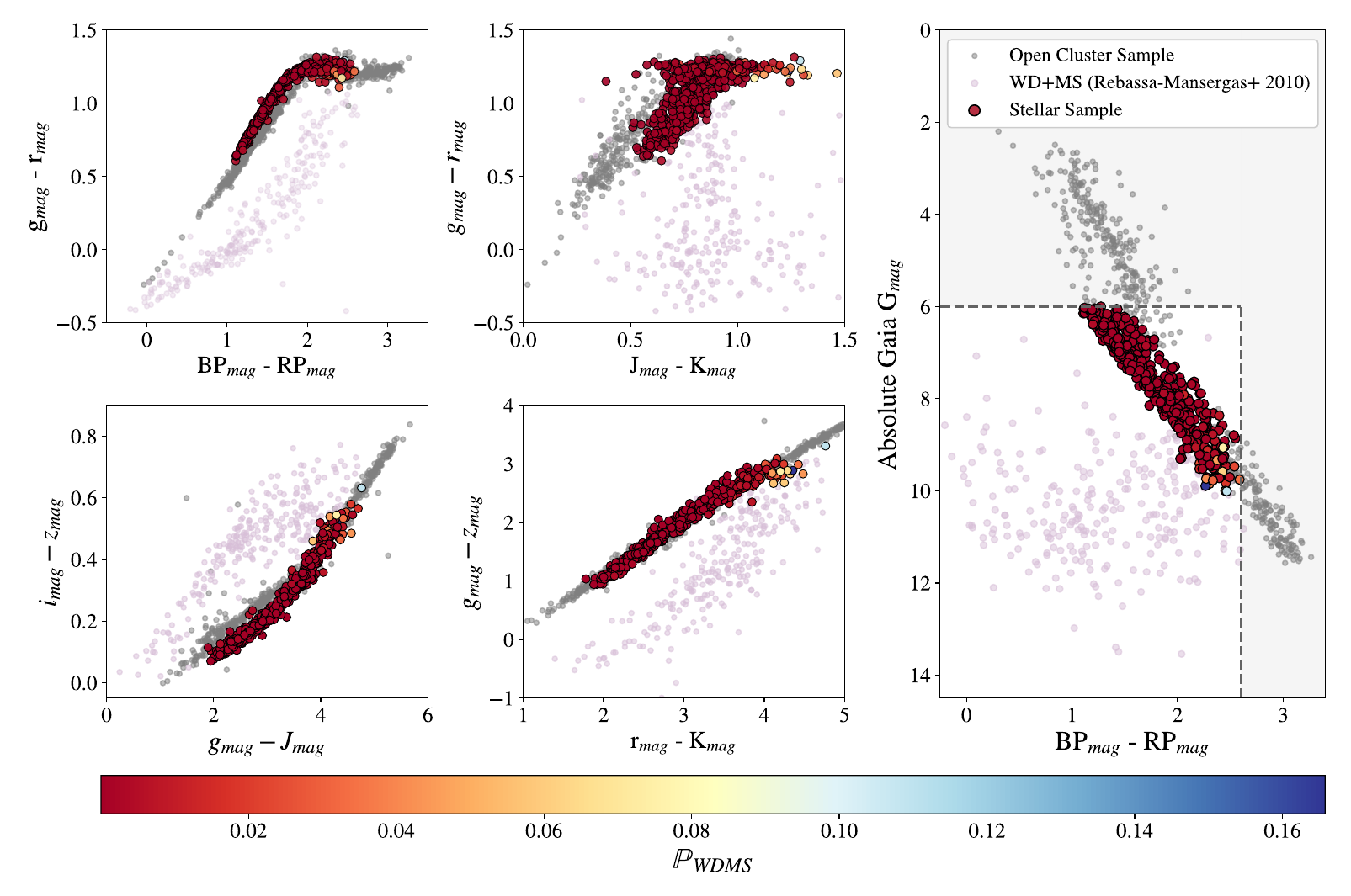}
    \caption{NGC 2682: A cluster with no identified high-probability or mid-probability candidate WD+MS binaries.}
    \label{fig:ngc2682-fullsample}
\end{figure*}

\begin{figure*}[!ht]
    \centering
    \includegraphics[width=0.62\textwidth]{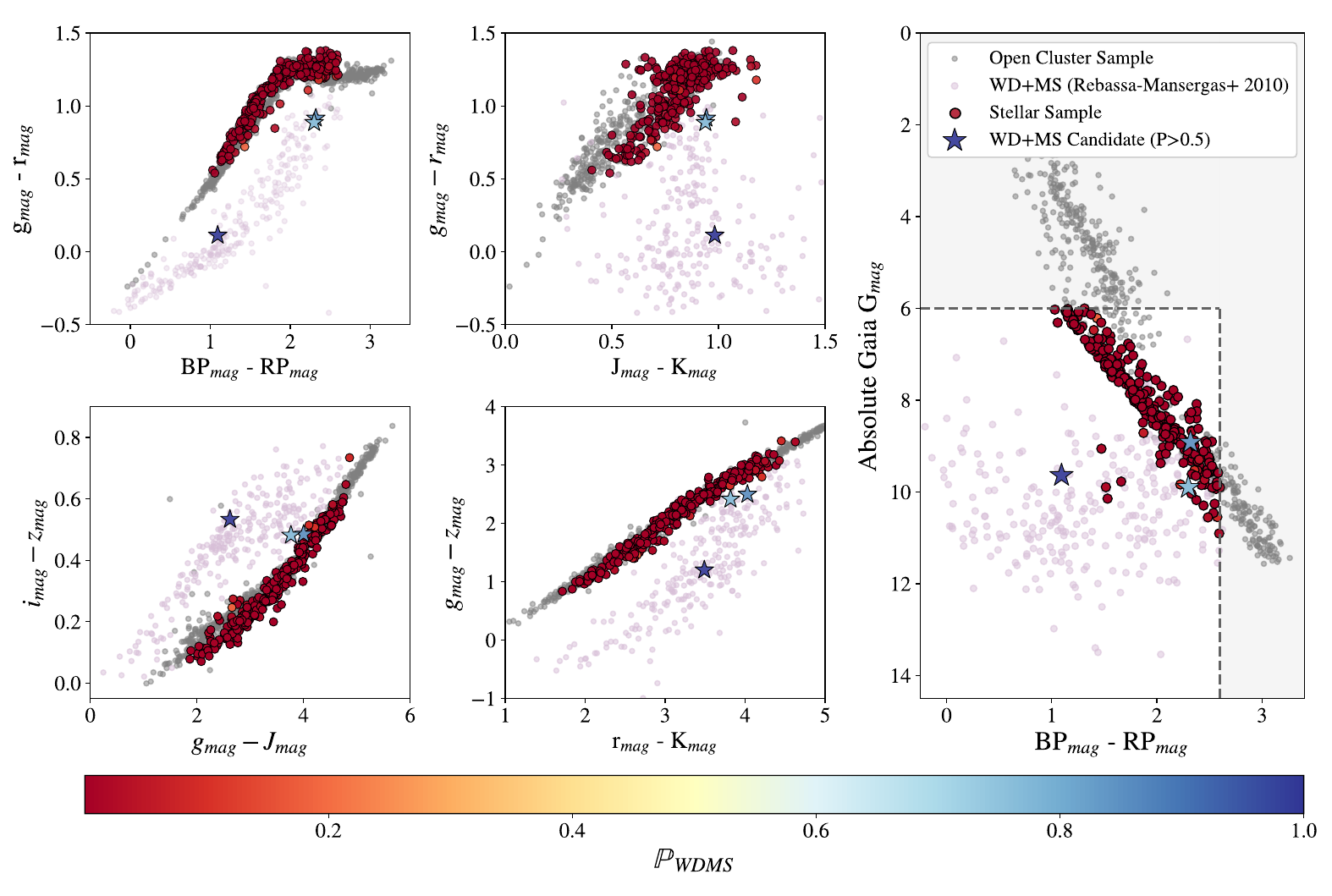}
    \caption{Alessi 12: A cluster with both high-probability and mid-probability candidate WD+MS binaries.}
    \label{fig:alessi12-fullsample}
\end{figure*}

Here, we present results from the SVM classification for two example OCs: Alessi 12 and NGC 2682. As seen in both Figures \ref{fig:ngc2682-fullsample} and \ref{fig:alessi12-fullsample}, the majority of sources in the Alessi 12 and NGC 2682 stellar samples lie along/near the ZAMS in both sample color-color spaces and a \textit{Gaia} CMD. While the SVM does not identify any probable WD+MS candidate binaries in NGC 2682, one high-probability source ($P>0.9$) and two medium-probability sources ($P>0.5$) are found in Alessi 12.

\newpage

\section{Photometric Errors of High-Probability Candidates}\label{ap:photo-errs}

\begin{figure*}[!ht]
    \centering
    \includegraphics[width=1\textwidth]{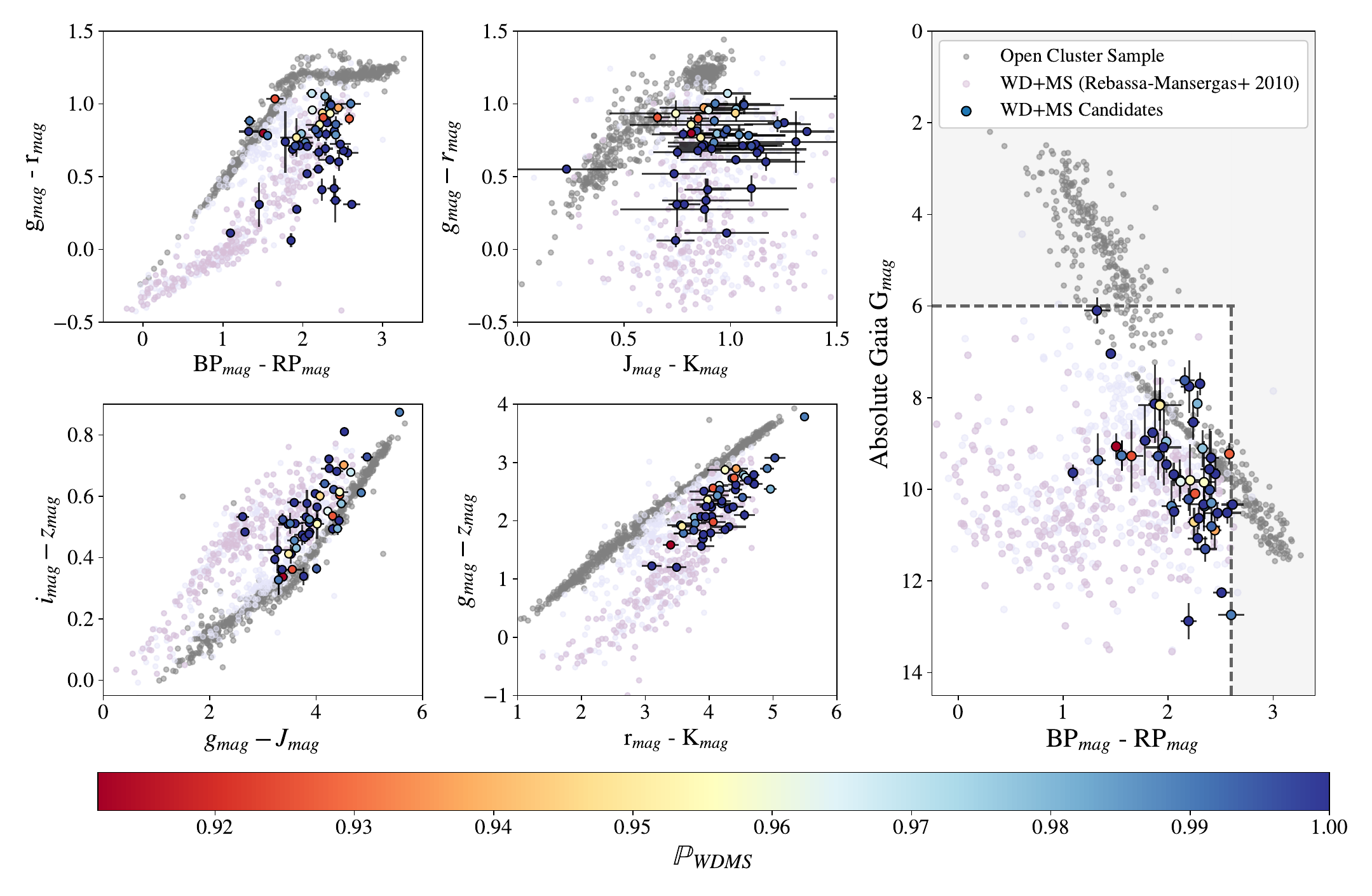}
    \caption{Photometric errors of the high-probability candidate WD+MS binaries in this catalog.}
    \label{fig:photerrors}
\end{figure*}

In Figure \ref{fig:photerrors}, we present the photometric errors of all 52 high-probability candidate WD+MS binaries in this catalog. Overall, the photometry (including error bars) spans the full WD+MS sample, with only small overlap with the ZAMS. We note that the largest photmetric errors arise from 2MASS $J$ and $K$ bands.

\section{Spectral Energy Distributions}\label{sec:seds}

As described in Section \ref{sec:additionalvetting}, we perform additional vetting of our high-probability candidates through a visual inspection of candidate SEDs. In Figures \ref{fig:allseds-1} and \ref{fig:allseds-2}, we present SEDs for all 52 high-probability candidate WD+MS systems in our catalog. We include both Pan-STARRS $g$, $r$, $i$, $z$ and $y$ band along with 2MASS $J$, $H$ and $K$ band photometry. To guide the eye, we have included five MS Pickles models \citep{1998PASP..110..863P} spanning K0V to M5V spectral types. However, we emphasize that the observational SEDs have not been corrected for Galactic extinction.

\begin{figure*}[!ht]
    \centering
    \includegraphics[width=0.95\textwidth]{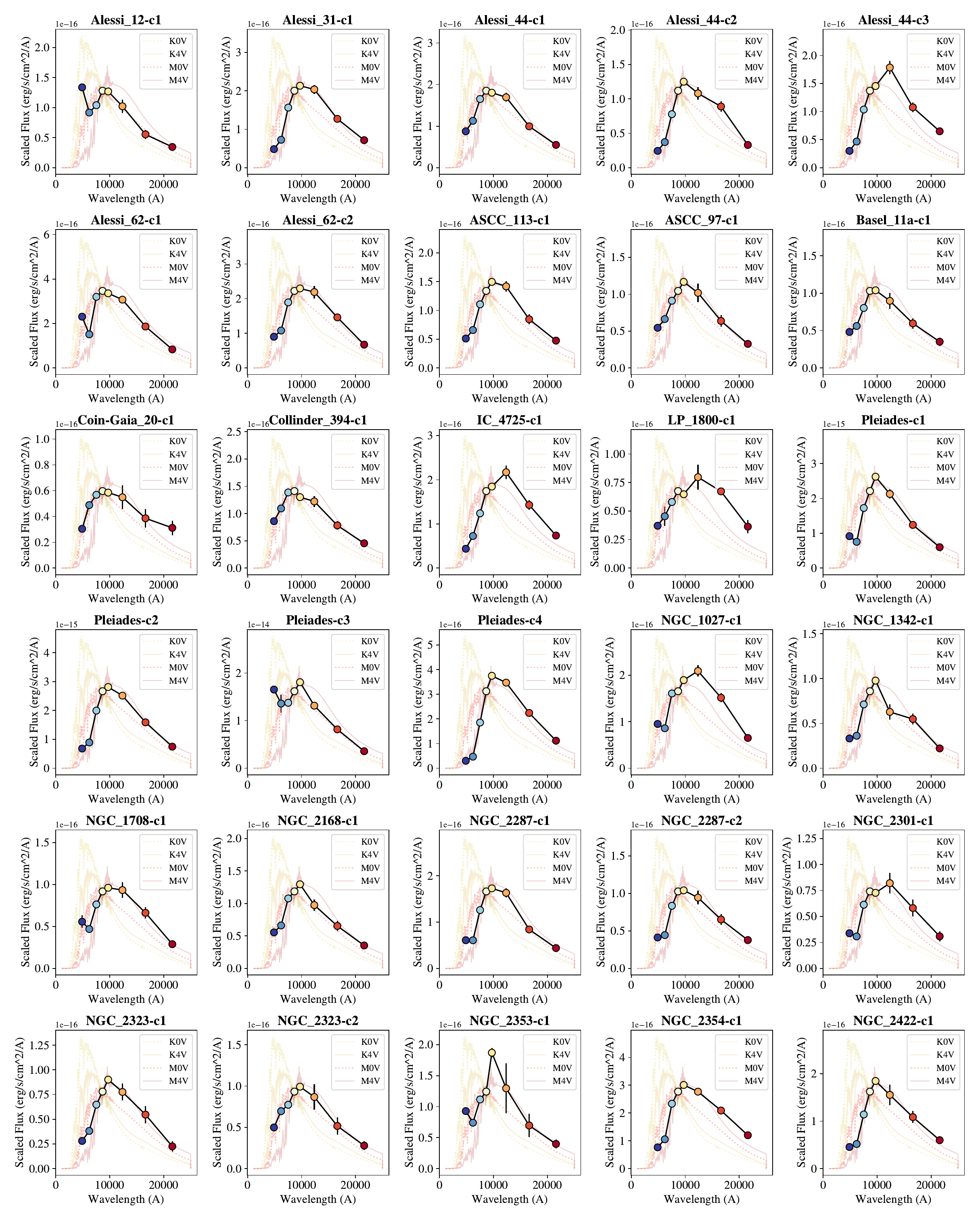}
    \caption{Spectral energy distributions of the high-probability candidate WD+MS binaries in this catalog.}
    \label{fig:allseds-1}
\end{figure*}

\begin{figure*}
    \centering
    \includegraphics[width=0.95\textwidth]{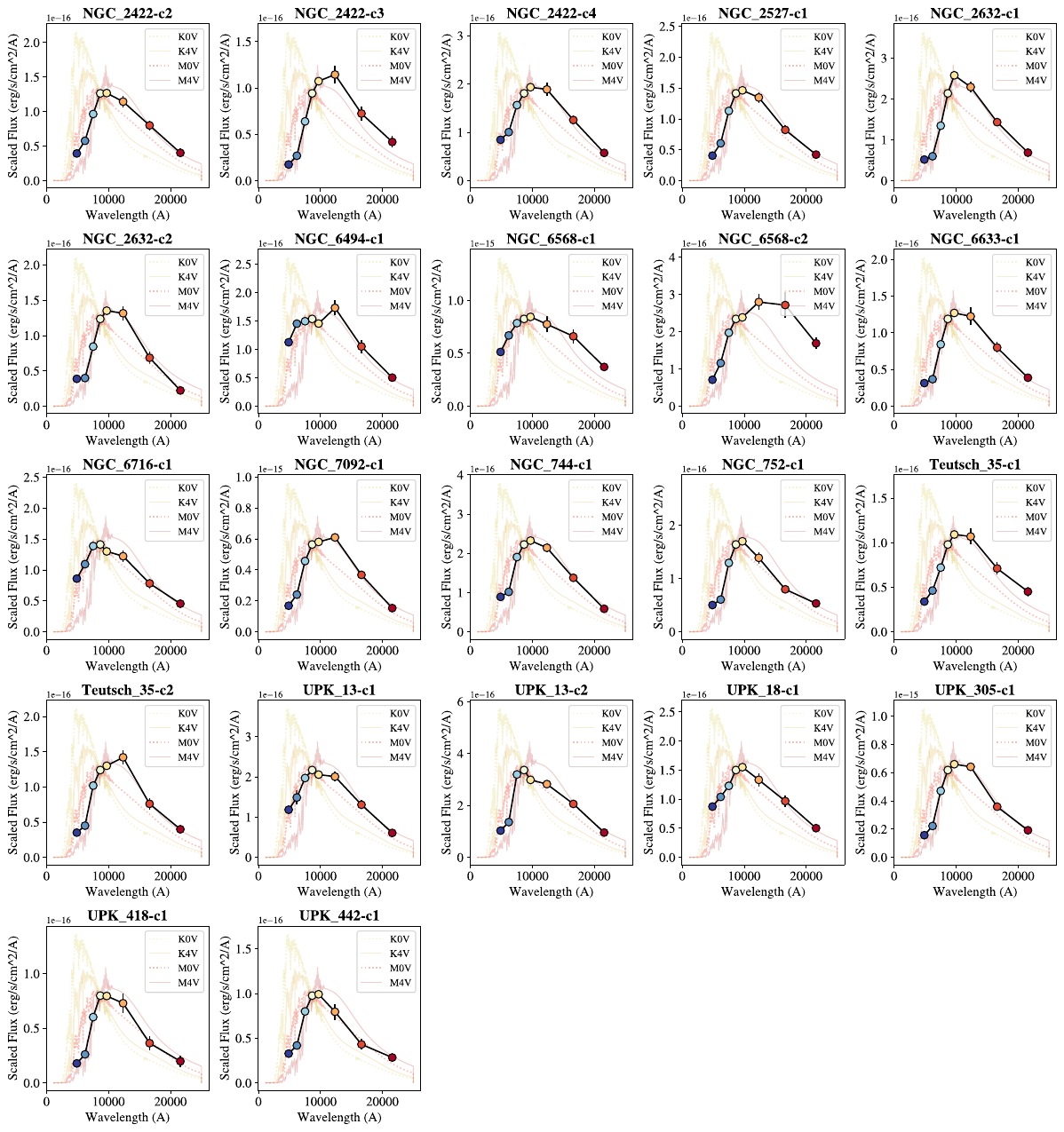}
    \caption{Continuation of Figure \ref{fig:allseds-1}.}
    \label{fig:allseds-2}
\end{figure*}

\end{appendix}
\end{document}